\message
{JNL.TEX version 0.95 as of 5/13/90.  Using CM fonts.}

\catcode`@=11
\expandafter\ifx\csname inp@t\endcsname\relax\let\inp@t=\input
\def\input#1 {\expandafter\ifx\csname #1IsLoaded\endcsname\relax
\inp@t#1%
\expandafter\def\csname #1IsLoaded\endcsname{(#1 was previously loaded)}
\else\message{\csname #1IsLoaded\endcsname}\fi}\fi
\catcode`@=12

\font\twelverm=cmr12			\font\twelvei=cmmi12
\font\twelvesy=cmsy10 scaled 1200	\font\twelveex=cmex10 scaled 1200
\font\twelvebf=cmbx12			\font\twelvesl=cmsl12
\font\twelvett=cmtt12			\font\twelveit=cmti12
\font\twelvesc=cmcsc10 scaled 1200	\font\twelvesf=cmss12
\font\twelvemib=cmr10 scaled 1200
                     
\font\tenmib=cmr10
\font\eightmib=cmr10 scaled 800 

\skewchar\twelvei='177			\skewchar\twelvesy='60
\skewchar\twelvemib='177

\newfam\mibfam

\def\twelvepoint{\normalbaselineskip=12.4pt plus 0.1pt minus 0.1pt
  \abovedisplayskip 12.4pt plus 3pt minus 9pt
  \belowdisplayskip 12.4pt plus 3pt minus 9pt
  \abovedisplayshortskip 0pt plus 3pt
  \belowdisplayshortskip 7.2pt plus 3pt minus 4pt
  \smallskipamount=3.6pt plus1.2pt minus1.2pt
  \medskipamount=7.2pt plus2.4pt minus2.4pt
  \bigskipamount=14.4pt plus4.8pt minus4.8pt
  \def\rm{\fam0\twelverm}          \def\it{\fam\itfam\twelveit}%
  \def\sl{\fam\slfam\twelvesl}     \def\bf{\fam\bffam\twelvebf}%
  \def\mit{\fam 1}                 \def\cal{\fam 2}%
  \def\sc{\twelvesc}		   \def\tt{\twelvett}%
  \def\sf{\twelvesf}               \def\mib{\fam\mibfam\twelvemib}%
  \textfont0=\twelverm   \scriptfont0=\tenrm   \scriptscriptfont0=\sevenrm
  \textfont1=\twelvei    \scriptfont1=\teni    \scriptscriptfont1=\seveni
  \textfont2=\twelvesy   \scriptfont2=\tensy   \scriptscriptfont2=\sevensy
  \textfont3=\twelveex   \scriptfont3=\twelveex\scriptscriptfont3=\twelveex
  \textfont\itfam=\twelveit
  \textfont\slfam=\twelvesl
  \textfont\bffam=\twelvebf \scriptfont\bffam=\tenbf
                            \scriptscriptfont\bffam=\sevenbf
  \textfont\mibfam=\twelvemib \scriptfont\mibfam=\tenmib
                              \scriptscriptfont\mibfam=\eightmib
  \normalbaselines\rm}


\mathchardef\alpha="710B
\mathchardef\beta="710C
\mathchardef\gamma="710D
\mathchardef\delta="710E
\mathchardef\epsilon="710F
\mathchardef\zeta="7110
\mathchardef\eta="7111
\mathchardef\theta="7112
\mathchardef\iota="7113
\mathchardef\kappa="7114
\mathchardef\lambda="7115
\mathchardef\mu="7116
\mathchardef\nu="7117
\mathchardef\xi="7118
\mathchardef\pi="7119
\mathchardef\rho="711A
\mathchardef\sigma="711B
\mathchardef\tau="711C
\mathchardef\phi="711E
\mathchardef\chi="711F
\mathchardef\psi="7120
\mathchardef\omega="7121
\mathchardef\varepsilon="7122
\mathchardef\vartheta="7123
\mathchardef\varpi="7124
\mathchardef\varrho="7125
\mathchardef\varsigma="7126
\mathchardef\varphi="7127


\def\beginlinemode{\endmode
  \begingroup\parskip=0pt \obeylines\def\\{\par}\def\endmode{\par\endgroup}}
\def\beginparmode{\endmode
  \begingroup \def\endmode{\par\endgroup}}
\let\endmode=\par
{\obeylines\gdef\
{}}
\def\singlespace{\baselineskip=\normalbaselineskip}

\def\oneandahalfspace{\baselineskip=\normalbaselineskip
  \multiply\baselineskip by 3 \divide\baselineskip by 2}
\def\doublespace{\baselineskip=\normalbaselineskip \multiply\baselineskip by 2}

\newcount\firstpageno
\firstpageno=2
\footline={\ifnum\pageno<\firstpageno{\hfil}\else{\hfil\twelverm\folio\hfil}\fi}
\def\toppageno{\global\footline={\hfil}\global\headline
  ={\ifnum\pageno<\firstpageno{\hfil}\else{\hfil\twelverm\folio\hfil}\fi}}
\let\rawfootnote=\footnote		
\def\footnote#1#2{{\rm\singlespace\parindent=0pt\parskip=0pt
  \rawfootnote{#1}{#2\hfill\vrule height 0pt depth 6pt width 0pt}}}
\def\raggedcenter{\leftskip=4em plus 12em \rightskip=\leftskip
  \parindent=0pt \parfillskip=0pt \spaceskip=.3333em \xspaceskip=.5em
  \pretolerance=9999 \tolerance=9999
  \hyphenpenalty=9999 \exhyphenpenalty=9999 }
\def\dateline{\rightline{\ifcase\month\or
  January\or February\or March\or April\or May\or June\or
  July\or August\or September\or October\or November\or December\fi
  \space\number\year}}
\def\received{\vskip 3pt plus 0.2fill
 \centerline{\sl (Received\space\ifcase\month\or
  January\or February\or March\or April\or May\or June\or
  July\or August\or September\or October\or November\or December\fi
  \qquad, \number\year)}}


\hsize=6.5truein
\hoffset=0pt
\vsize=8.9truein
\voffset=0pt
\parskip=\medskipamount
\def\\{\cr}
\twelvepoint		
\doublespace		
\overfullrule=0pt	




\def\title			
  {\null\vskip 3pt plus 0.2fill
   \beginlinemode \doublespace \raggedcenter \bf}

\def\author			
  {\vskip 3pt plus 0.2fill \beginlinemode
   \singlespace \raggedcenter\sc}

\def\affil			
  {\vskip 3pt plus 0.1fill \beginlinemode
   \oneandahalfspace \raggedcenter \sl}

\def\abstract			
  {\vskip 3pt plus 0.3fill \beginparmode
   \oneandahalfspace ABSTRACT: }

\def\endtitlepage		
  {\endpage			
   \body}
\let\endtopmatter=\endtitlepage

\def\body			
  {\beginparmode}		

\def\head#1{			
  \goodbreak\vskip 0.5truein	
  {\immediate\write16{#1}
   \raggedcenter \uppercase{#1}\par}
   \nobreak\vskip 0.25truein\nobreak}

\def\subhead#1{			
  \vskip 0.25truein		
  {\raggedcenter {#1} \par}
   \nobreak\vskip 0.25truein\nobreak}

\def\beginitems{
\par\medskip\bgroup\def\i##1 {\item{##1}}\def\ii##1 {\itemitem{##1}}
\leftskip=36pt\parskip=0pt}
\def\enditems{\par\egroup}

\def\beneathrel#1\under#2{\mathrel{\mathop{#2}\limits_{#1}}}

\def\refto#1{$^{#1}$}		

\def\references			
  {\head{References}		
   \beginparmode
   \frenchspacing \parindent=0pt \leftskip=1truecm
   \parskip=8pt plus 3pt \everypar{\hangindent=\parindent}}

\gdef\refis#1{\item{#1.\ }}			

\gdef\journal#1, #2, #3, 1#4#5#6{		
    {\sl #1~}{\bf #2}, #3 (1#4#5#6)}		

\def\endreferences{\body}

\def\figurecaptions		
  {\endpage
   \beginparmode
   \head{Figure Captions}
}

\def\endpage			
  {\vfill\eject}

\def\endpaper			
  {\endmode\vfill\supereject}


\def\heading				
  {\vskip 0.5truein plus 0.1truein	
   \beginparmode \def\\{\par} \parskip=0pt \singlespace \raggedcenter}

\def\subheading				
  {\vskip 0.25truein plus 0.1truein	
   \beginlinemode \singlespace \parskip=0pt \def\\{\par}\raggedcenter}

\def\tag#1$${\eqno(#1)$$}

\def\align#1$${\eqalign{#1}$$}

\def\aligntag#1$${\gdef\tag##1\\{&(##1)\cr}\eqalignno{#1\\}$$
  \gdef\tag##1$${\eqno(##1)$$}}

\def\endaligntag{}

\def\overset #1\to#2{{\mathop{#2}\limits^{#1}}}
\def\underset#1\to#2{{\let\next=#1\mathpalette\undersetpalette#2}}
\def\undersetpalette#1#2{\vtop{\baselineskip0pt
\ialign{$\mathsurround=0pt #1\hfil##\hfil$\crcr#2\crcr\next\crcr}}}


\def\ref#1{Ref.~#1}			
\def\Ref#1{Ref.~#1}			
\def\[#1]{[\cite{#1}]}
\def\cite#1{{#1}}
\let\eq=\Eq\let\eqs=\Eqs		
\def\(#1){(\call{#1})}
\def\call#1{{#1}}
\def\taghead#1{}
\def\frac#1#2{{#1 \over #2}}

\def\12{{1\over2}}

\def\etal{{\it et al.\ }}

\def\sla{\raise.15ex\hbox{$/$}\kern-.57em}
\def\leaderfill{\leaders\hbox to 1em{\hss.\hss}\hfill}
\def\twiddle{\lower.9ex\rlap{$\kern-.1em\scriptstyle\sim$}}
\def\bigtwiddle{\lower1.ex\rlap{$\sim$}}
\def\gtwid{\mathrel{\raise.3ex\hbox{$>$\kern-.75em\lower1ex\hbox{$\sim$}}}}
\def\ltwid{\mathrel{\raise.3ex\hbox{$<$\kern-.75em\lower1ex\hbox{$\sim$}}}}
\def\square{\kern1pt\vbox{\hrule height 1.2pt\hbox{\vrule width 1.2pt\hskip 3pt
   \vbox{\vskip 6pt}\hskip 3pt\vrule width 0.6pt}\hrule height 0.6pt}\kern1pt}
\def\tdot#1{\mathord{\mathop{#1}\limits^{\kern2pt\ldots}}}

\def\pmb#1{\setbox0=\hbox{#1}%
  \kern-.025em\copy0\kern-\wd0
  \kern  .05em\copy0\kern-\wd0
  \kern-.025em\raise.0433em\box0 }

\catcode`@=11
\newcount\r@fcount \r@fcount=0
\newcount\r@fcurr
\immediate\newwrite\reffile
\newif\ifr@ffile\r@ffilefalse
\def\w@rnwrite#1{\ifr@ffile\immediate\write\reffile{#1}\fi\message{#1}}

\def\writer@f#1>>{}
\def\referencefile{
  \r@ffiletrue\immediate\openout\reffile=\jobname.ref%
  \def\writer@f##1>>{\ifr@ffile\immediate\write\reffile%
    {\noexpand\refis{##1} = \csname r@fnum##1\endcsname = %
     \expandafter\expandafter\expandafter\strip@t\expandafter%
     \meaning\csname r@ftext\csname r@fnum##1\endcsname\endcsname}\fi}%
  \def\strip@t##1>>{}}

\def\citeall#1{\xdef#1##1{#1{\noexpand\cite{##1}}}}
\def\cite#1{\each@rg\citer@nge{#1}}	

\def\each@rg#1#2{{\let\thecsname=#1\expandafter\first@rg#2,\end,}}
\def\first@rg#1,{\thecsname{#1}\apply@rg}	
\def\apply@rg#1,{\ifx\end#1\let\next=\relax
\else,\thecsname{#1}\let\next=\apply@rg\fi\next}

\def\citer@nge#1{\citedor@nge#1-\end-}	
\def\citer@ngeat#1\end-{#1}
\def\citedor@nge#1-#2-{\ifx\end#2\r@featspace#1 
  \else\citel@@p{#1}{#2}\citer@ngeat\fi}	
\def\citel@@p#1#2{\ifnum#1>#2{\errmessage{Reference range #1-#2\space is bad.}%
    \errhelp{If you cite a series of references by the notation M-N, then M and
    N must be integers, and N must be greater than or equal to M.}}\else%
 {\count0=#1\count1=#2\advance\count1 by1\relax\expandafter\r@fcite\the\count0,%
  \loop\advance\count0 by1\relax
    \ifnum\count0<\count1,\expandafter\r@fcite\the\count0,%
  \repeat}\fi}

\def\r@featspace#1#2 {\r@fcite#1#2,}	
\def\r@fcite#1,{\ifuncit@d{#1}
    \newr@f{#1}%
    \expandafter\gdef\csname r@ftext\number\r@fcount\endcsname%
                     {\message{Reference #1 to be supplied.}%
                      \writer@f#1>>#1 to be supplied.\par}%
 \fi%
 \csname r@fnum#1\endcsname}
\def\ifuncit@d#1{\expandafter\ifx\csname r@fnum#1\endcsname\relax}%
\def\newr@f#1{\global\advance\r@fcount by1%
    \expandafter\xdef\csname r@fnum#1\endcsname{\number\r@fcount}}

\let\r@fis=\refis			
\def\refis#1#2#3\par{\ifuncit@d{#1}
   \newr@f{#1}%
   \w@rnwrite{Reference #1=\number\r@fcount\space is not cited up to now.}\fi%
  \expandafter\gdef\csname r@ftext\csname r@fnum#1\endcsname\endcsname%
  {\writer@f#1>>#2#3\par}}

\def\ignoreuncited{
   \def\refis##1##2##3\par{\ifuncit@d{##1}%
     \else\expandafter\gdef\csname r@ftext\csname r@fnum##1\endcsname\endcsname%
     {\writer@f##1>>##2##3\par}\fi}}

\def\r@ferr{\endreferences\errmessage{I was expecting to see
\noexpand\endreferences before now;  I have inserted it here.}}
\let\r@ferences=\references
\def\references{\r@ferences\def\endmode{\r@ferr\par\endgroup}}

\let\endr@ferences=\endreferences
\def\endreferences{\r@fcurr=0
  {\loop\ifnum\r@fcurr<\r@fcount
    \advance\r@fcurr by 1\relax\expandafter\r@fis\expandafter{\number\r@fcurr}%
    \csname r@ftext\number\r@fcurr\endcsname%
  \repeat}\gdef\r@ferr{}\endr@ferences}


\let\r@fend=\endpaper\gdef\endpaper{\ifr@ffile
\immediate\write16{Cross References written on []\jobname.REF.}\fi\r@fend}

\catcode`@=12

\citeall\refto		
\citeall\ref		%
\citeall\Ref		%

\catcode`@=11
\newcount\tagnumber\tagnumber=0

\immediate\newwrite\eqnfile
\newif\if@qnfile\@qnfilefalse
\def\write@qn#1{}
\def\writenew@qn#1{}
\def\w@rnwrite#1{\write@qn{#1}\message{#1}}
\def\@rrwrite#1{\write@qn{#1}\errmessage{#1}}

\def\taghead#1{\gdef\t@ghead{#1}\global\tagnumber=0}
\def\t@ghead{}

\expandafter\def\csname @qnnum-3\endcsname
  {{\t@ghead\advance\tagnumber by -3\relax\number\tagnumber}}
\expandafter\def\csname @qnnum-2\endcsname
  {{\t@ghead\advance\tagnumber by -2\relax\number\tagnumber}}
\expandafter\def\csname @qnnum-1\endcsname
  {{\t@ghead\advance\tagnumber by -1\relax\number\tagnumber}}
\expandafter\def\csname @qnnum0\endcsname
  {\t@ghead\number\tagnumber}
\expandafter\def\csname @qnnum+1\endcsname
  {{\t@ghead\advance\tagnumber by 1\relax\number\tagnumber}}
\expandafter\def\csname @qnnum+2\endcsname
  {{\t@ghead\advance\tagnumber by 2\relax\number\tagnumber}}
\expandafter\def\csname @qnnum+3\endcsname
  {{\t@ghead\advance\tagnumber by 3\relax\number\tagnumber}}

\def\equationfile{%
  \@qnfiletrue\immediate\openout\eqnfile=\jobname.eqn%
  \def\write@qn##1{\if@qnfile\immediate\write\eqnfile{##1}\fi}
  \def\writenew@qn##1{\if@qnfile\immediate\write\eqnfile
    {\noexpand\tag{##1} = (\t@ghead\number\tagnumber)}\fi}
}

\def\callall#1{\xdef#1##1{#1{\noexpand\call{##1}}}}
\def\call#1{\each@rg\callr@nge{#1}}

\def\each@rg#1#2{{\let\thecsname=#1\expandafter\first@rg#2,\end,}}
\def\first@rg#1,{\thecsname{#1}\apply@rg}
\def\apply@rg#1,{\ifx\end#1\let\next=\relax%
\else,\thecsname{#1}\let\next=\apply@rg\fi\next}

\def\callr@nge#1{\calldor@nge#1-\end-}
\def\callr@ngeat#1\end-{#1}
\def\calldor@nge#1-#2-{\ifx\end#2\@qneatspace#1 %
  \else\calll@@p{#1}{#2}\callr@ngeat\fi}
\def\calll@@p#1#2{\ifnum#1>#2{\@rrwrite{Equation range #1-#2\space is bad.}
\errhelp{If you call a series of equations by the notation M-N, then M and
N must be integers, and N must be greater than or equal to M.}}\else%
 {\count0=#1\count1=#2\advance\count1 by1\relax\expandafter\@qncall\the\count0,%
  \loop\advance\count0 by1\relax%
    \ifnum\count0<\count1,\expandafter\@qncall\the\count0,%
  \repeat}\fi}

\def\@qneatspace#1#2 {\@qncall#1#2,}
\def\@qncall#1,{\ifunc@lled{#1}{\def\next{#1}\ifx\next\empty\else
  \w@rnwrite{Equation number \noexpand\(>>#1<<) has not been defined yet.}
  >>#1<<\fi}\else\csname @qnnum#1\endcsname\fi}

\let\eqnono=\eqno
\def\eqno(#1){\tag#1}
\def\tag#1$${\eqnono(\displayt@g#1 )$$}

\def\aligntag#1\endaligntag
  $${\gdef\tag##1\\{&(##1 )\cr}\eqalignno{#1\\}$$
  \gdef\tag##1$${\eqnono(\displayt@g##1 )$$}}

\def\eqalignno#1{\displ@y \tabskip\centering
  \halign to\displaywidth{\hfil$\displaystyle{##}$\tabskip\z@skip
    &$\displaystyle{{}##}$\hfil\tabskip\centering
    &\llap{$\displayt@gpar##$}\tabskip\z@skip\crcr
    #1\crcr}}

\def\displayt@gpar(#1){(\displayt@g#1 )}

\def\displayt@g#1 {\rm\ifunc@lled{#1}\global\advance\tagnumber by1
        {\def\next{#1}\ifx\next\empty\else\expandafter
        \xdef\csname @qnnum#1\endcsname{\t@ghead\number\tagnumber}\fi}%
  \writenew@qn{#1}\t@ghead\number\tagnumber\else
        {\edef\next{\t@ghead\number\tagnumber}%
        \expandafter\ifx\csname @qnnum#1\endcsname\next\else
        \w@rnwrite{Equation \noexpand\tag{#1} is a duplicate number.}\fi}%
  \csname @qnnum#1\endcsname\fi}

\def\ifunc@lled#1{\expandafter\ifx\csname @qnnum#1\endcsname\relax}

\let\@qnend=\end\gdef\end{\if@qnfile
\immediate\write16{Equation numbers written on []\jobname.EQN.}\fi\@qnend}

\catcode`@=12

\newread\epsffilein    
\newif\ifepsffileok    
\newif\ifepsfbbfound   
\newif\ifepsfverbose   
\newdimen\epsfxsize    
\newdimen\epsfysize    
\newdimen\epsftsize    
\newdimen\epsfrsize    
\newdimen\epsftmp      
\newdimen\pspoints     
\pspoints=1bp          
\epsfxsize=0pt         
\epsfysize=0pt         
\def\epsfbox#1{\global\def\epsfllx{72}\global\def\epsflly{72}%
   \global\def\epsfurx{540}\global\def\epsfury{720}%
   \def\lbracket{[}\def\testit{#1}\ifx\testit\lbracket
   \let\next=\epsfgetlitbb\else\let\next=\epsfnormal\fi\next{#1}}%
\def\epsfgetlitbb#1#2 #3 #4 #5]#6{\epsfgrab #2 #3 #4 #5 .\\%
   \epsfsetgraph{#6}}%
\def\epsfnormal#1{\epsfgetbb{#1}\epsfsetgraph{#1}}%
\def\epsfgetbb#1{%
%
%
\openin\epsffilein=#1
\ifeof\epsffilein\errmessage{I couldn't open #1, will ignore it}\else
%
%
   {\epsffileoktrue \chardef\other=12
    \def\do##1{\catcode`##1=\other}\dospecials \catcode`\ =10
    \loop
       \read\epsffilein to \epsffileline
       \ifeof\epsffilein\epsffileokfalse\else
%
%
          \expandafter\epsfaux\epsffileline:. \\%
       \fi
   \ifepsffileok\repeat
   \ifepsfbbfound\else
    \ifepsfverbose\message{No bounding box comment in #1; using defaults}\fi\fi
   }\closein\epsffilein\fi}%
%
%
\def\epsfclipstring{}
\def\epsfsetgraph#1{%
   \epsfrsize=\epsfury\pspoints
   \advance\epsfrsize by-\epsflly\pspoints
   \epsftsize=\epsfurx\pspoints
   \advance\epsftsize by-\epsfllx\pspoints
%
%
   \epsfxsize\epsfsize\epsftsize\epsfrsize
   \ifnum\epsfxsize=0 \ifnum\epsfysize=0
      \epsfxsize=\epsftsize \epsfysize=\epsfrsize
      \epsfrsize=0pt
%
%
     \else\epsftmp=\epsftsize \divide\epsftmp\epsfrsize
       \epsfxsize=\epsfysize \multiply\epsfxsize\epsftmp
       \multiply\epsftmp\epsfrsize \advance\epsftsize-\epsftmp
       \epsftmp=\epsfysize
       \loop \advance\epsftsize\epsftsize \divide\epsftmp 2
       \ifnum\epsftmp>0
          \ifnum\epsftsize<\epsfrsize\else
             \advance\epsftsize-\epsfrsize \advance\epsfxsize\epsftmp \fi
       \repeat
       \epsfrsize=0pt
     \fi
   \else \ifnum\epsfysize=0
     \epsftmp=\epsfrsize \divide\epsftmp\epsftsize
     \epsfysize=\epsfxsize \multiply\epsfysize\epsftmp   
     \multiply\epsftmp\epsftsize \advance\epsfrsize-\epsftmp
     \epsftmp=\epsfxsize
     \loop \advance\epsfrsize\epsfrsize \divide\epsftmp 2
     \ifnum\epsftmp>0
        \ifnum\epsfrsize<\epsftsize\else
           \advance\epsfrsize-\epsftsize \advance\epsfysize\epsftmp \fi
     \repeat
     \epsfrsize=0pt
    \else
     \epsfrsize=\epsfysize
    \fi
   \fi
%
%
   \ifepsfverbose\message{#1: width=\the\epsfxsize, height=\the\epsfysize}\fi
   \epsftmp=10\epsfxsize \divide\epsftmp\pspoints
   \vbox to\epsfysize{\vfil\hbox to\epsfxsize{%
      \ifnum\epsfrsize=0\relax
        \includegraphics{#1}%
      \else
        \epsfrsize=10\epsfysize \divide\epsfrsize\pspoints
        \includegraphics{#1}%
      \fi
      \hfil}}%
\global\epsfxsize=0pt\global\epsfysize=0pt}%
%
%
{\catcode`\%=12 \global\let\epsfpercent=
%
%
\long\def\epsfaux#1#2:#3\\{\ifx#1\epsfpercent
   \def\testit{#2}\ifx\testit\epsfbblit
      \epsfgrab #3 . . . \\%
      \epsffileokfalse
      \global\epsfbbfoundtrue
   \fi\else\ifx#1\par\else\epsffileokfalse\fi\fi}%
%
%
\def\epsfempty{}%
\def\epsfgrab #1 #2 #3 #4 #5\\{%
\global\def\epsfllx{#1}\ifx\epsfllx\epsfempty
      \epsfgrab #2 #3 #4 #5 .\\\else
   \global\def\epsflly{#2}%
   \global\def\epsfurx{#3}\global\def\epsfury{#4}\fi}%
%
%
\def\epsfsize#1#2{\epsfxsize}
%
%

\def\Kbar{{\overline{K}}}

{\parindent=0pt February 1997 \hfill{Wash. U. HEP/97-60} }
\rightline{hep-lat/9702015}

\title Chiral perturbation theory for $K^+ \rightarrow \pi^+ \pi^0$ decay in the
 continuum and on the lattice

\author Maarten F.L. Golterman${}^{1}$%
\footnote{}{${}^{1}$ e-mail: maarten@aapje.wustl.edu}%
\ and Ka Chun Leung${}^{2}$%
\footnote{}{${}^{2}$ e-mail: leung@hbar.wustl.edu}%
\affil Department of Physics 
       Washington University 
       St. Louis, MO 63130, USA

\abstract{In this paper we use one-loop chiral perturbation theory in order to
compare lattice computations of the $K^+\to\pi^+\pi^0$ decay amplitude with the
experimental value.  This makes it possible to investigate three systematic
effects that plague lattice computations: quenching, finite-volume effects, and
the fact that lattice computations have been done at unphysical values of the
quark masses and pion external momenta (only this latter effect shows up at
tree level).  We apply our results to the most recent lattice computation (ref. 
[\cite{lattdecay3}]), and find that all three effects are substantial.  We
conclude that one-loop corrections in chiral perturbation theory help in
explaining the discrepancy between lattice results and the real-world value. 
We also revisit $B_K$, which is closely related to the $K^+\to\pi^+\pi^0$ decay 
amplitude by chiral symmetry.}

\endtopmatter

\subhead{\bf 1. Introduction}

One of the aims of practitioners of lattice QCD is the accurate computation of
QCD weak matrix elements.  There are two reasons for this.  First, such matrix
elements are needed in order to explain the experimental values of certain
quantities.  A prominent example is the $\Delta I=1/2$ rule for nonleptonic kaon
decay. Second, precise knowledge of the values of weak matrix elements helps 
with
the determination of less well-known entries in the Cabibbo--Kobayashi--Maskawa
matrix.  An important example is the kaon B-parameter $B_K$, which parametrizes
the strength of CP-violation in $K^0-\Kbar^0$ mixing. (For these and various 
other phenomenological aspects of kaon weak interactions, see ref. 
[\cite{phreviews}] and references therein; for the lattice approach, see refs. 
[\cite{BandSreview,Claudetasi,sharpetasi}] and references therein.)

Actual lattice calculations 
[\cite{lattBK0,lattBK,lattdecay1,lattdecay2,lattdecay3}]
are hampered by many systematic errors.  Even if the continuum limit is 
obtained, several important sources of error still remain.  Two of these are
caused by the use of a finite volume and by the use of the quenched
approximation.  A third systematic effect typically occurs for matrix elements
involving more than two external particles.  Taking $K\to 2\pi$ decay as an
example, the matrix element most easily determined from a numerical computation
is that with degenerate quark masses, and
with all external particles (both the kaon and the two pions in this
case) at rest.  This does not correspond to the physical situation, where of
course $m_K>2m_\pi$, and
each of the final pions has a spatial momentum $|{\bf
k}|=\sqrt{{1\over  4}m_K^2-m_\pi^2}$
(in the kaon's rest frame). This implies that such lattice results will have to
be extrapolated to physical values of the quark masses and the external momenta.

All of the above described effects can be studied in chiral perturbation theory
(ChPT). Finite-volume effects can be calculated in ChPT following ref.
[\cite{fv}], while a quenched formulation of ChPT (``quenched chiral 
perturbation
theory," or QChPT) was developed in ref. [\cite{them1,sharpe0}]. The third systematic
effect introduced above can also be investigated in ChPT by simply calculating
both the ``unphysical" matrix element (with all external particles at rest) and
the ``physical" one (with the correct kinematics).  The ratio of the two then
gives an estimate of the factor needed in order to convert the
unphysical matrix element into the physical one. This was done at tree level for
the decay $K^+\to\pi^+\pi^0$ in refs. [\cite{lattdecay2,lattdecay3}] (at tree
level no finite-volume or quenching effects show up in ChPT). It was found that
even with the conversion factor (which also corrects for the fact that the quark
masses in numerical computations are different from the
physical quark masses), the lattice value of the $K^+\to\pi^+\pi^0$ matrix
element is about two times larger than its experimental
value.

In this paper we study the $K^+\to\pi^+\pi^0$ weak matrix element and $B_K$ in
ChPT to one-loop order.  This is interesting for various reasons.  Nontrivial
finite-volume and quenching effects show up in ChPT first at one loop, and
therefore a one-loop calculation will give us insight into the size of these
effects.  Also, there are one-loop corrections to the tree-level conversion
factor between unphysical and physical $K^+\to\pi^+\pi^0$ matrix elements,
and one would like to know whether these corrections can be part of the
explanation that numerical results appear to come out too large
[\cite{Claudetasi}]. Moreover, it has been suggested that one can expect such
one-loop effects to be sizeable, since they include an estimate of the effect of
final-state interactions of the two pions [\cite{isguretal}].

We restrict ourselves to $B_K$ and $K^+\to\pi^+\pi^0$ decay because these two
quantities require the introduction of only one new parameter at the lowest 
order in
ChPT. This is because the corresponding weak operators are both components 
of
the same $SU(3)$-flavor 27-plet, and this 27-plet has a unique representation
in  lowest order ChPT. It would of course be interesting to also
consider $K^0\to \pi^+\pi^-$ decay, but this would lead to the introduction of
new parameters (corresponding to the $SU(3)$ octet operators that contribute to
this decay).  Also, it has been observed that inclusion of the $\sigma$ 
resonance
may be required for a full understanding of this decay on the lattice 
[\cite{sigresonance}], which would
take us outside of the systematic approach of ChPT.

Beyond lowest order in ChPT, a large number of new operators exist which can
contribute to $B_K$ and to $K^+\to\pi^+\pi^0$ decay [\cite{Kambor1}], each with
its accompanying new parameter (the so-called $O(p^4)$ low-energy constants). 
The
values of these new parameters are largely unknown.  Therefore, at one loop we
can only estimate the size of the nonanalytic terms (using a reasonable value of
the cutoff), because the nonanalytic terms do not depend on these new 
parameters.
In this paper, we will assume that using a reasonable value of the cutoff in the
nonanalytic terms and setting the $O(p^4)$ low-energy constants to zero will 
lead to a 
valid
estimate of the size of corrections to tree-level ChPT results. Some idea about
the quantitative uncertainties introduced by this approach can be obtained from
the sensitivity of our results to the value of the cutoff. These issues will all 
be
discussed in more detail in due course in this paper.

Let us end the introduction with an overview of the paper.  We revisit $B_K$ in
section 3 (quenching and finite-volume effects for $B_K$ have been calculated
within ChPT before in refs. [\cite{sharpe0,sharpetasi}]), while the calculations 
for
the $K^+\to\pi^+\pi^0$ matrix element are contained in sections 4--6. In section
3 we also discuss our approach to the use of ChPT at one loop in some more
detail. In section 4 we present our calculation of the nonanalytic one-loop
corrections for the physical matrix element.  Section 5 explains which Euclidean
correlation function is computed in numerical lattice QCD, and in section 6 we 
present in
some detail the calculation of this correlation function in one-loop (Q)ChPT. 
For
this calculation we restrict ourselves to the case of degenerate quark masses
($m_u=m_d=m_s$). In section 7, we use the results of the previous sections in
order to obtain quantitative estimates of the various systematic effects for
typical lattice computations. We start with calculating the size of the
nonanalytic one-loop corrections in the real world, as was done before by 
Bijnens
\etal in ref. [\cite{Bijnens}] and by various other authors 
[\cite{bardeen,Kambor2,Kambor3,Bruno}]. We
then estimate the one-loop correction to the conversion of the unphysical 
lattice
$K^+\to\pi^+\pi^0$ matrix element to the physical one in the unquenched theory.
Next, we include the changes that result from using the quenched approximation 
for the
unphysical matrix element, and we discuss the numerical results of ref. 
[\cite{lattdecay3}].
The uncertainties introduced by our lack of knowledge of all low-energy 
constants
are discussed. We end this section by considering a quantity which is basically
the ratio of $B_K$ and the $K^+\to\pi^+\pi^0$ matrix element [\cite{treerel}]. 
This quantity is
interesting because it is independent of the tree-level low-energy constant, and
therefore not sensitive to ambiguities in its value.  All
calculations are repeated for the finite spatial volume which was used in ref.
[\cite{lattdecay3}]. Our conclusions are in section 8. Section 2 contains a
summary of other (Q)ChPT results that we will need in this paper. Appendix A
presents an adaptation of the quenched $B_K$ result relevant to staggered fermion 
QCD, and
Appendix B relates our result for the $K^+\to\pi^+\pi^0$ correlation function to
the general analysis of such Euclidean correlation functions in ref.
[\cite{MandT}].  Appendix C deals with a technicality.

\subhead{\bf 2. Essentials of ChPT and QChPT}

We start with a summary of results in ChPT [\cite{wein,powercount,Gasser}]
that we will need in this paper. This also gives us an opportunity to 
establish our notation. We will refer to the theory without quenching as the
``full" or ``unquenched" theory. 
The $SU(3)$-octet Goldstone meson fields are organized in the nonlinear 
representation
$$\Sigma =\exp \left({2i\phi \over f}\right)\; , \eqno(sigma)$$
where $\phi$ is the hermitian matrix
$$\phi =\pmatrix{
{{\pi^0 \over \sqrt{2}}+{\eta \over \sqrt{6}}}&\pi^+&K^+\cr
   \pi^-&{-{\pi^0 \over \sqrt{2}} +{\eta \over \sqrt{6}}}&K^0\cr
   K^-&\Kbar^0&-{2\eta \over \sqrt{6}}\cr
}\ ,\eqno(field)$$
and $f$ is a parameter with the dimension of a mass.
Masses of the $u$, $d$ and $s$ quarks are 
incorporated in the mass matrix
$$M=\pmatrix{m_u&0&0\cr 0&m_d&0\cr 0&0&m_s}\ . \eqno(mass)$$
To lowest order (which we will refer to as $O(p^2)$) in ChPT, the 
Euclidean Lagrangian is constructed 
with the introduction of an additional parameter $v$:
$${\cal L}={f^2 \over 8}{\rm Tr} \Bigl( \partial^\mu \Sigma \partial_\mu 
\Sigma^\dagger \Bigr) -v{\rm Tr} \Bigl( M\Sigma^\dagger +M^\dagger 
\Sigma \Bigr)\ . \eqno(L)$$
One can read off the tree-level meson masses from the quadratic terms,
after expanding $\Sigma$ in $\phi$ 
(for simplicity we will choose $m_u =m_d =m$):
$$m^2_{\pi ,0}={ {8{\it v}m} \over f^2}\ \ ,\ \ \ m^2_{K,0}
={ {4{\it v}(m +m_s)} \over f^2}\ \ ,\ \ \ m^2_{\eta ,0}={ {8{\it v}(m +2m_s)} 
\over {3f^2}}\ \ . \eqno(treemass)$$
It follows that
$$m^2_{\eta ,0}={4\over3}m^2_{K,0}-{1\over3}m^2_{\pi ,0}\ , \eqno(massrelation)$$
which predicts a value of the $\eta$-mass about 3\% too large.

At tree level, the parameter $f$ is the weak decay constant,
$f=f_\pi =f_K$. At one loop 
the pion and kaon decay constants $f_\pi$ and $f_K$ and corresponding
wavefunction renormalizations $Z_\pi$ and $Z_K$ are given by
$$\eqalignno{
f_\pi &=f\ \Biggl(1+{m^2_K \over (4\pi f)^2}\left[ I_\pi +K_1 +K_2 \; y_\pi 
\right]\Biggr)\ , &(fpifull)\cr
Z_\pi &=1-{2m^2_K \over {3(4\pi f)^2}}\left[ I_\pi +K_1 +K_2 \; y_\pi \right]
&(Zpifull)
}$$
for the pion and
$$\eqalignno{
f_K &=f\ \Biggl(1+{m^2_K \over (4\pi f)^2}\left[ I_K +\Bigl({{K_1 \over 2}+
K_2}\Bigr) +{K_1 \over 2} y_\pi \right]\Biggr)\ , &(fKfull)\cr
Z_K &=1-{2m^2_K \over {3(4\pi f)^2}}\left[ I_K +\Bigl({{K_1 \over 2}+K_2}\Bigr)
+{K_1 \over 2} y_\pi \right]
&(ZKfull)
}$$
for the kaon, where 
$$\eqalignno{
I_\pi &=-\log{m^2_K \over m_\pi^2} -\Bigl({1+2y_\pi}\Bigr) \log{m^2_\pi 
\over \Lambda^2}\ , &(Ipi)\cr
I_K &=-{3\over 2}\log {m^2_K \over m^2_\pi} -\Bigl({1-{1\over 4}y_\pi}\bigr)
\log {m^2_\eta \over m^2_\pi} -{1\over 2}\Bigl({5+y_\pi}\Bigr)
\log {m^2_\pi \over \Lambda^2}\ . &(IK)\cr
}$$
In these equations, we made use of the fact that we can replace the tree-level 
quantities $m^2_{\pi ,0}$, $m^2_{K,0}$ and $m^2_{\eta ,0}$ by their renormalized 
physical values $m^2_\pi$, $m^2_K$ and $m^2_\eta$. The parameter $y_\pi$ is defined as
$$y_\pi = {m^2_\pi \over m^2_K}\ . \eqno(ypi)$$
$\Lambda$ is the cutoff, and the dimensionless parameters $K_1$ and 
$K_2$ are contact-term coefficients coming from the $O(p^4)$ chiral
Lagrangian [\cite{Gasser}].

A Lagrangian formulation of quenched ChPT (QChPT) was developed in ref.
[\cite{them1}]. The basic feature is the introduction of (unphysical) 
spin $1/2$ ``ghost" quarks with bosonic statistics and the same masses as the
physical quarks, so that their virtual quark-loop 
contributions cancel those from the original quarks
[\cite{morel}]. As a result, the 
$\eta'$ meson is light [\cite{sharpe00,them1,sharpe0}], and therefore needs to 
be
kept in the chiral Lagrangian. The original octet \eq{sigma} has to be 
extended into a nonet to include the $\eta'$:
$$\Sigma_q = e^{i\eta' / \scriptscriptstyle{\sqrt{3}}f_q}\ \Sigma \ , 
\eqno(sigmaq)$$
where $f_q$ is the quenched parameter equivalent to $f$ in the full theory, and 
$f$ is replaced by $f_q$ in \eq{sigma}.
In the mass-degenerate case $m_u =m_d =m_s$, the $\eta'$ two-point 
function acquires both single- and double-pole terms:
$$G_{\eta' ,\eta'} (p,q)\equiv \langle {\eta'}(p){\eta'}(q)\rangle =\delta (p+q) 
\Biggl( {1 \over {p^2 +m^2_{\pi ,0}}} -{{\mu^2 +\alpha p^2}\over 
{(p^2 +m^2_{\pi ,0})}^2}\Biggr)\ ,\eqno(Getaeta)$$
where $\mu^2$ is a parameter which would correspond to the singlet part of 
the $\eta'$ mass in full QCD. In full QCD, an estimate using the experimental 
mass of the $\eta'$ gives: ${\mu^2}/3 \approx {(500\; {\rm MeV})}^2$ 
(for $\alpha =0$). In the mass-nondegenerate case $m_u =m_d \neq m_s$, the
$\eta$ two-point function also acquires a double-pole term through mixing 
with the $\eta'$. It is then simpler to work in the (nondiagonal) basis of 
${\bar u}u,\ 
{\bar d}d,$ and ${\bar s}s$ meson states in the neutral sector
(labeled by $i=u,d,s$ respectively). Their two-point functions are:
$$G_{i,j} (p,q)=\delta (p+q)\left( {\delta_{ij} \over {p^2 +M^2_i}}-{{(\mu^2 
+\alpha p^2)/3} 
\over {(p^2 +M^2_i)(p^2 +M^2_j)}}\right)\ ,\eqno(generalGs) $$
where $M^2_i \equiv 8{\it v}m_i/f^2_q$. When $m_u =m_d \neq m_s$, it is easy 
to see that $M^2_u =M^2_d =m^2_{\pi ,0}$ and $M^2_s =2m^2_{K,0} -m^2_{\pi ,0}$. We define
$$\delta \equiv{{\mu^2 /3}\over {8\pi^2 f^2_q}}\eqno(delta)$$
for later use.

The quenched one-loop expressions for the weak decay constants $f_\pi$ and 
$f_K$ and wavefunction renormalizations $Z_\pi$ and $Z_K$ in the 
mass-nondegenerate case $(m_u =m_d \neq m_s )$ are
$$\eqalignno{
f_\pi &=f_q\;\Biggl( 1+{m^2_\pi \over {(4\pi f_q)}^2}{\widetilde K}\Biggr)\ , 
&(fpiquenched)\cr
Z_\pi &=1-{2m^2_\pi \over 3{(4\pi f_q)}^2}{\widetilde K}\ , &(Zpiquenched)\cr
f_K &=f_q\;\Biggl( 1+{\widetilde I}_K +{m^2_K \over (4\pi f_q)^2}
{\widetilde K}\Biggr)\ , &(fKquenched)\cr
Z_K &=1-{2\over 3}\left[ {\widetilde I}_K +{m^2_K \over (4\pi f_q)^2}
{\widetilde K}\right]\ ,&(ZKquenched)
}$$
where 
$${\widetilde I}_K =-{1\over 2}\delta \left( 1+{1\over {2(1-y_\pi)}}
\log{\left( {y_\pi \over {2-y_\pi}}\right)}\right) +{1\over 3}
\alpha {m^2_K \over (4\pi f_q)^2}\left( 1+{{y_\pi (2-y_\pi )} \over 
{2(1-y_\pi )}}\log{\left( {y_\pi \over {2-y_\pi}}\right)}\right), 
\eqno(wtIK)$$
and the dimensionless parameter ${\widetilde K}$ is again a contact-term 
coefficient from the $O(p^4 )$ Lagrangian. In contrast to the 
full theory, there is only one such coefficient in the quenched case. 
In ChPT, an $O(p^4)$ operator containing the trace of the mass matrix $M$, 
\eq{mass}, when evaluated at tree level, contributes to a particular linear 
combination of $K_1$ and $K_2$ in the unquenched quantities. However, in 
QChPT, the same operator is converted into one where the 
supertrace [\cite{them1}] of the mass 
matrix, now enlarged to include the ghost-quark masses, is taken, and hence its
contribution to \eqs{fpiquenched,Zpiquenched,fKquenched,ZKquenched} vanishes. 
Note that chiral logarithms and $\delta$ or $\alpha$ dependence are absent 
in \eq{fpiquenched} and \eq{Zpiquenched}.  This is a consequence of
taking $m_u=m_d$.

In general, the double poles in \eq{generalGs} lead to severe infra-red 
divergences in physical quantities in QChPT. In fact, physical quantities 
with a singular chiral limit and/or anomalous volume dependence were found 
in refs. [\cite{them1,sharpe0,relation,sharpe1,fvscatt,Booth,zhang}]. An example 
is $f_K$ in 
\eq{fKquenched}, which diverges in the limit $y_\pi\to 0$.

\subhead{\bf 3. Review of $B_K$}

The kaon B-parameter $B_K$ is defined as
$$B_K={{\langle {\overline K}^0 \vert ({\bar s}d{\bar s}d)_{LL} 
\vert K^0 \rangle} \over {8 \over 3}{f_K^2 m_K^2}}\ , \eqno(BK)$$
in which the four-quark operator is defined by
$$({\bar q_1} q_2 {\bar q_3} q_4)_{LL} =({\bar q_{1L}} \gamma^\mu q_{2L})
({\bar q_{3L}} \gamma_\mu q_{4L})\ , \eqno(anycurcur)$$
where $q_L={1 \over 2}(1-\gamma_5)q$ is a left-handed quark field. The 
denominator in \eq{BK} is the matrix element 
${\langle {\overline K}^0 \vert ({\bar s}d{\bar s}d)_{LL} 
\vert K^0 \rangle}$ evaluated by vacuum saturation. 
In ChPT, weak-interaction operators, the chiral transformation properties of 
which are dictated by the Standard Model, are constructed from the Goldstone 
meson 
field $\Sigma$ and the mass matrix $M$. The operator 
$({\bar s}d{\bar s}d)_{LL}$ is a component of a 27-plet of $SU(3)_L$ and has
$\Delta S=2$. To lowest order in ChPT $(O(p^2))$, there is only one 27-plet
operator: 
$$O'=\alpha_{\scriptscriptstyle 27} \; t^{ij}_{kl}({\Sigma \partial^\mu 
\Sigma^\dagger})_{ik}({\Sigma \partial_\mu \Sigma^\dagger})_{jl} \ , 
\eqno(OOprime)$$
where the tensor $t^{ij}_{kl} $ satisfies:
$$\sum_i t^{ij}_{ik}=0\ ,\ \ \sum_i t^{ij}_{ki}=0\ \ {\rm and}\  \ t^{ij}_{kl} 
=t^{ji}_{lk}\ . \eqno(Oprime)$$
In order to select the $\Delta S=2$ component, we set 
$$t^{22}_{33}=1\eqno(Oprimets)$$
and all other components to zero. The parameter $\alpha_{\scriptscriptstyle 27}$ 
is not constrained by any symmetry and its value is determined by QCD dynamics.
At tree level (or order $p^2$), one obtains 
$$\langle {\overline K}^0 |O'|K^0 \rangle={8\alpha_{\scriptscriptstyle 27} 
\over f^2}m^2_K\; . \eqno(Oprimetree)$$
The factor $m^2_K$ comes from a factor $p^2_K$ where we take the external 
kaon momentum $p_K$ on shell, {\it i.e.} $m_K$ here is the renormalized physical 
kaon mass. Therefore,
$$B_K^{full} = {3\alpha_{\scriptscriptstyle 27} \over f^4} \equiv B^{f}
\eqno(B)$$
($f_K =f$ at tree level).

In this paper, we are interested in corrections of order one higher in ChPT 
power counting [\cite{powercount,Gasser}]. We will refer to these as $O(p^4)$ 
contributions. In general, they consist of one-loop corrections coming from 
the $O(p^2)$ operators along with tree-level contributions from $O(p^4)$ 
operators. A large number of $O(p^4)$ operators with the symmetries of
the operator in \eq{BK} can be constructed from $\Sigma$ and $M$; see ref. 
[\cite{Kambor1}] for a complete list of $O(p^4)$ operators that 
are relevant to strangeness-changing nonleptonic weak-decay processes involving
up to four Goldstone mesons. Each of them is associated with an arbitrary 
coefficient, which, just like $\alpha_{\scriptscriptstyle 27}$, 
cannot be determined within ChPT. Since at $O(p^4)$ only tree-level 
contributions
from these  operators are needed, contact terms with arbitrary coefficients 
result. In contrast, one-loop contributions from $O(p^2)$ operators generically
give rise to nonanalytic functions of the quark masses with coefficients
determined by the $O(p^2)$ operators, and a cutoff $\Lambda$ naturally arises. 
The dependence on the cutoff of the $O(p^4)$-operator coefficients can
then be chosen such that physical quantities are independent of the cutoff. 

In order to determine the $O(p^4)$-operator coefficients, many authors have used 
combinations of particle phenomenology and theoretical 
ideas such as large-$N_c$ arguments, resonance-dominance models, etc.
(see for instance refs. 
[\cite{bardeen,Kambor2,Kambor3,Bruno,Gasser,resonance,Pichetal}]). A less 
phenomenological approach is to find relations 
between different physical quantities that are independent of $O(p^2)$- and 
$O(p^4)$-operator coefficients. An illustrative example is the
combination $f_\eta f^{1/3}_\pi / f^{4/3}_K$ of 
the pion, kaon and eta decay constants [\cite{Gasser}]; 
examples in QChPT can be found in 
refs. [\cite{relation,zakopane,zhang}]. 
In this approach, physical predictions at order $p^4$ are made without the 
model-dependent uncertainties mentioned above. These relations of course
are precisely the Ward-identities associated with spontaneously broken chiral 
symmetry. Theoretically, an infinite number of such relations can be 
constructed. However, the large number of low-energy
constants at order $p^4$ makes this difficult in practice. 
This is certainly true when we only consider $B_K$ and 
$K^+ \rightarrow \pi^+ \pi^0$ decay, both of which depend on a large
number of $O(p^4)$ low-energy constants.
(A program to determine some of these coefficients from experiment
in the $m_\pi =0$ 
approximation has been carried out in refs. [\cite{Kambor2,Kambor3}].) 
In this paper, we will limit ourselves mainly to the nonanalytic one-loop 
corrections. With a reasonable choice of the cutoff, the magnitude of the 
corrections to the tree-level expressions for certain
quantities can be estimated.

Next, we present the results for the full and quenched $K^0-\Kbar^0$
matrix elements of the operator $O'$ to one loop. For the full
case, we obtain
$$\eqalignno{
&{\langle {\overline K}^0 |O'|K^0 \rangle}^f =
{8\alpha_{\scriptscriptstyle 27}m^2_K \over f^2}\Biggl( 1+{m^2_K \over 
(4\pi f)^2}\biggl[ I
+F_1 +F_2 +F_3 -(F_2 +2F_3 )\; y_\pi +F_3 \; y^2_\pi \cr
&\phantom{
{\langle {\overline K}^0 |O'|K^0 \rangle}^f =
{8\alpha_{\scriptscriptstyle 27}m^2_K \over f^2}\Biggl( 1+{m^2_K \over 
(4\pi f_\pi )^2}\biggl[ I
}
-{2\over 3}\left( {K_1 \over 2}+K_2 \right)-{K_1\over 3}\; y_\pi \biggr]
\Biggr)\ , &(1lfOprime)
}$$
where
$$I=-5\log{m^2_K\over m^2_\pi}-{1\over 6}\Bigl (4-y_\pi \Bigr) 
\Bigl( 10-y_\pi \Bigr)\log {m^2_\eta \over m^2_\pi }-{1\over 3}
\Bigl(35-y_\pi +2y^2_\pi\Bigr)\log{m^2_\pi \over \Lambda^2}\ ,\eqno(I)$$
while for the quenched case
$$\eqalignno{
&{\langle {\overline K}^0 |O'|K^0 \rangle}^q ={8\alpha^q_{\scriptscriptstyle 
27}m^2_K \over f^2_q}\Biggl( 1+{m^2_K 
\over (4\pi f_q)^2}\biggl[ \; {\widetilde I}+{\widetilde F}_1 
+{\widetilde F}_3 -2{\widetilde F}_3 \; y_\pi +{\widetilde F}_3 
\; y^2_\pi -{2{\widetilde K}\over 3}\biggr]\cr
&\phantom{
{\langle {\overline K}^0 |O'|K^0 \rangle}^q ={8\alpha^q_{\scriptscriptstyle 
27}m^2_K \over f^2_q}\Biggl( 1
}
+\delta\; I_\delta +{2\over 3}
\alpha {m^2_K \over (4\pi f_q)^2}I_\alpha \Biggr)\ , &(1lqOprime)
}$$
where
$$\eqalignno{
{\widetilde I}&=-2(4-2y_\pi +y^2_\pi )\log{m^2_K\over \Lambda^2} -y_\pi 
(1+y_\pi )\log{y_\pi} -(2-y_\pi )(3-y_\pi )\log{(2-y_\pi )}\ ,&(I1)\cr
I_\delta &=-3-{{2+2y_\pi -y^2_\pi}\over 2(1-y_\pi )}
\log{\left( y_\pi \over {2-y_\pi}\right)}\ ,&(I2)\cr
I_\alpha &=2(1-y_\pi )^2 \log{\biggl( {m^2_\pi \over \Lambda^2} \biggr)}
-{{2-y_\pi}\over {2(1-y_\pi)}}\biggl( 2-7y_\pi +2y^2_\pi \biggr)
\log{\left( y_\pi \over {2-y_\pi}\right)}+4-2y_\pi +y^2_\pi \ ,
\phantom{1111}
&(I3)\cr
}$$
with $y_\pi$ defined in \eq{ypi} and $\delta$ in \eq{delta}. 
In QChPT, the operator $({\bar s}d{\bar s}d)_{LL}$ is represented by 
\eq{OOprime} with $\alpha_{\scriptscriptstyle 27}$ replaced by a new 
parameter $\alpha^q_{\scriptscriptstyle 27}$. 
The mass relation \eq{massrelation} has been used to eliminate $m_\eta$ in 
favor of $m_\pi$ and $m_K$, except in the $\eta$ chiral logarithm
of \eq{1lfOprime}.
The dimensionless parameters $F_i$ and ${\widetilde F}_j$ 
with $i=1,2,3$ and $j=1,3$ come 
from $O(p^4)$ weak operators. They are parametrized in such a way as to 
obtain simple expressions for $B_K$ below. 
There are just two instead of three independent contact-term coefficients in 
the quenched case: the reason is the same as that for the analogous
difference between the full and quenched $O(p^4)$ expressions
for $f_{\pi,K}$ and $Z_{\pi,K}$ in the previous section. These results do not
include finite-volume corrections, which are exponentially small in 
$m_\pi L$ for large volume (where $L$ is the linear dimension 
of the spatial volume) [\cite{sharpe0}].

The results for $B_K$ to one loop for respectively the full and quenched 
theories in the mass-nondegenerate case are
$$\eqalignno{
B^{f}_K=B^{f}\Biggl( 1+{m^2_K \over {(4\pi f)}^2}\biggl[& 
-\left( 1-{3\epsilon \over 2}+{\epsilon^2 \over 2}\right) \log{m^2_\pi 
\over \Lambda^2}-2\log{m^2_K \over \Lambda^2}-3\left( 1+{\epsilon \over 2}
+{\epsilon^2 \over 18}\right) \log{m^2_\eta \over \Lambda^2}\cr
&-{8\over 3}\left(K_1 +K_2 \right) +{4K_1 \over 3}\epsilon +F_1 +F_2 
\epsilon +F_3 \epsilon^2 \biggr]\Biggr)
\ ,&(Bfull)
}$$
and
$$\eqalignno{
&B^{q}_K=B^{q}\times\cr
&\Biggl(1+{m^2_K \over (4\pi f_q)^2}\left[-2(3+ \epsilon ^2)
\log{m^2_K \over \Lambda^2}-(2+\epsilon ^2)\log(1-\epsilon ^2)-3\epsilon 
\log{{1+\epsilon}\over{1-\epsilon}}-{8{\widetilde K}\over 3}
+{\widetilde F}_1 +{\widetilde F}_3 
\epsilon^2 \right]\cr
&\phantom{
\Biggl(1
}
-\delta \left[-{{2-\epsilon ^2} \over {2\epsilon}}\log{{1+\epsilon} 
\over {1-\epsilon}}+2\right]\cr
&\phantom{
\Biggl(1
}
+{2 \over 3}\alpha {m^2_K \over (4\pi f_q)^2}\left[2\epsilon ^2
\log{\Biggl({m^2_K \over \Lambda ^2}(1-\epsilon)\Biggr)}-
{{1-2\epsilon ^2-\epsilon ^3} \over \epsilon}\log{{1+\epsilon} \over 
{1-\epsilon}}+2+\epsilon ^2\right]\Biggr)
\ .&(Bquenched)}$$
The parameter $\epsilon$ is defined as
$$\epsilon\equiv 1-{m^2_\pi \over m^2_K}={{m_s-m} 
\over {m_s+m}}=1-y_\pi \ .\eqno(ep)$$
(In Appendix A we present a modified expression for $B^{q}_K$ which may be 
applicable when 
quenched staggered fermions are used at nonzero lattice 
spacing and flavor symmetry is not fully
restored.) The parameter $B^{f}$ is defined in \eq{B}, and $B^q$ is defined from 
$\alpha^q_{\scriptscriptstyle 27}$ and $f_q$ by an
equation similar to \eq{B}. As is clear from
these expressions, the full and quenched theories are entirely different
theories. In particular, there is no reason 
to believe that the coefficients of
the $O(p^2)$ and $O(p^4)$ operators should be the same in ChPT and QChPT. The 
$\delta$ and $\alpha$ terms in 
\eq{Bquenched} come from the double poles in \eq{generalGs}. The full ChPT 
result \eq{Bfull} has previously been calculated in ref. [\cite{Bijnens}] in the $m_\pi =0$ 
approximation and in refs. [\cite{bardeen1,Kambor3,sharpe1,Bruno}]; 
whereas the QChPT 
result \eq{Bquenched} has previously been calculated in ref. [\cite{sharpe0}] for the
mass-nondegenerate case with $\alpha =\delta =0$, and in ref. [\cite{sharpetasi}] for the
mass-nondegenerate case with $\alpha=0$. We note a sign difference in 
the $\delta$ term between \eq{Bquenched} and the result in ref.
[\cite{sharpetasi}].

Note that, apart from the usual nonanalytic cutoff dependence,
nonanalytic functions depending on $\epsilon$ appear in 
\eq{Bquenched}.  This dependence is uniquely determined by (Q)ChPT, and
predicts the dependence of $B_K$ on $m_\pi$ and $m_K$ once the values
of the low-energy constants $K_i$, $\widetilde K$, $F_i$ and ${\widetilde F}_j$ have been
determined at a given value of the cutoff $\Lambda$ 
from lattice results for $B_K$ at given fixed values of these
masses.
Note that the contributions proportional to $\delta$ and $\alpha$ in 
$B_K^q$ vanish in the mass-degenerate case, $\epsilon=0$.  (This is
why we have not absorbed contact terms proportional to $\alpha$ into
${\widetilde F}_1$ and ${\widetilde F}_3$.)
We see that, apart from a change from $B^f$ to $B^q$ and $f$ to $f_q$, quenching 
does not
introduce any change in the nonanalytic one-loop corrections of $B_K$ in the
mass-degenerate case. As we will see, such is not the case for 
the $K^+ \rightarrow \pi ^+ \pi ^0$ matrix element. 

\subhead{\bf 4. $K^+ \rightarrow \pi^+ \pi^0$ in the real world}

The $\Delta S=1,\ \Delta I=3/2\ \; K^+ \rightarrow \pi^+ \pi^0$ decay amplitude 
is 
proportional to the weak matrix element 
$$\langle  \pi^+ \pi^0 |{\left( {\bar s}d{\bar u}u+{\bar s}u{\bar u}d-
{\bar s}d{\bar d}d\right)}_{LL} |K^+ \rangle \; . \eqno(currentcurrent)$$
The operator is the $\Delta I=3/2$ component of the same 27-plet that
also contains the operator ${({\bar s}d{\bar s}d)}_{LL}$ (\eq{BK}). To lowest 
order in ChPT, the $\Delta I=3/2\ \; O(p^2)$-operator is 
represented by
$$O_4=\alpha_{\scriptscriptstyle 27} \; r^{ij}_{kl}({\Sigma \partial^\mu 
\Sigma^\dagger})_{ik}({\Sigma \partial_\mu \Sigma^\dagger})_{jl}
\ , \eqno(rwOfour)$$
where the tensor $r^{ij}_{kl}$ has nonzero components 
$$r^{21}_{31}=r^{12}_{13}=r^{12}_{31}=r^{21}_{13}=
{1 \over 2}\ ,$$ 
$$r^{22}_{32}=r^{22}_{23}=-{1 \over 2}\eqno(ts)$$
(all other components vanish). It is easy to check that the tensor 
$r^{ij}_{kl}$ satisfies condition \eq{Oprime} . The parameter 
$\alpha_{\scriptscriptstyle 27}$ is the same as in \eq{OOprime}. This was 
used in ref. [\cite{treerel}] to derive a tree-level relation between the 
$K^+ \rightarrow \pi^+ \pi^0$ decay rate and $B_K$: at tree level, using \eq{B}, 
we have
$$\langle \pi^+ \pi^0 |O_4 |K^+ \rangle =
{{12i\alpha_{\scriptscriptstyle 27}} \over {\sqrt{2} f_\pi^3}}
\biggl( m^2_K -m^2_\pi \biggr)
={4i\over {\sqrt{2}\;}}f_\pi \left( m^2_K -m^2_\pi \right) B_K\ . 
\eqno(rwtreeOfour)$$
Again, the factor $m^2_K -m^2_\pi$ arises from taking the kaon and pion 
momenta on shell, analogous to \eq{Oprimetree}.

Bijnens \etal\ [\cite{Bijnens}] were the first to investigate $B_K$ and 
$K^+ \rightarrow \pi^+ \pi^0$ decay to one loop in ChPT. They used the 
approximation of vanishing pion mass based on the fact that, experimentally,
${m_\pi^2}/{m_K^2} \approx 1/13$.  Kambor \etal \ [\cite{Kambor2}]  
gave a general analysis of strangeness-changing nonleptonic weak-decay 
processes involving up to four Goldstone mesons including 
$K^+ \rightarrow \pi^+ \pi^0$, to one loop. 

\bigskip
\epsfbox{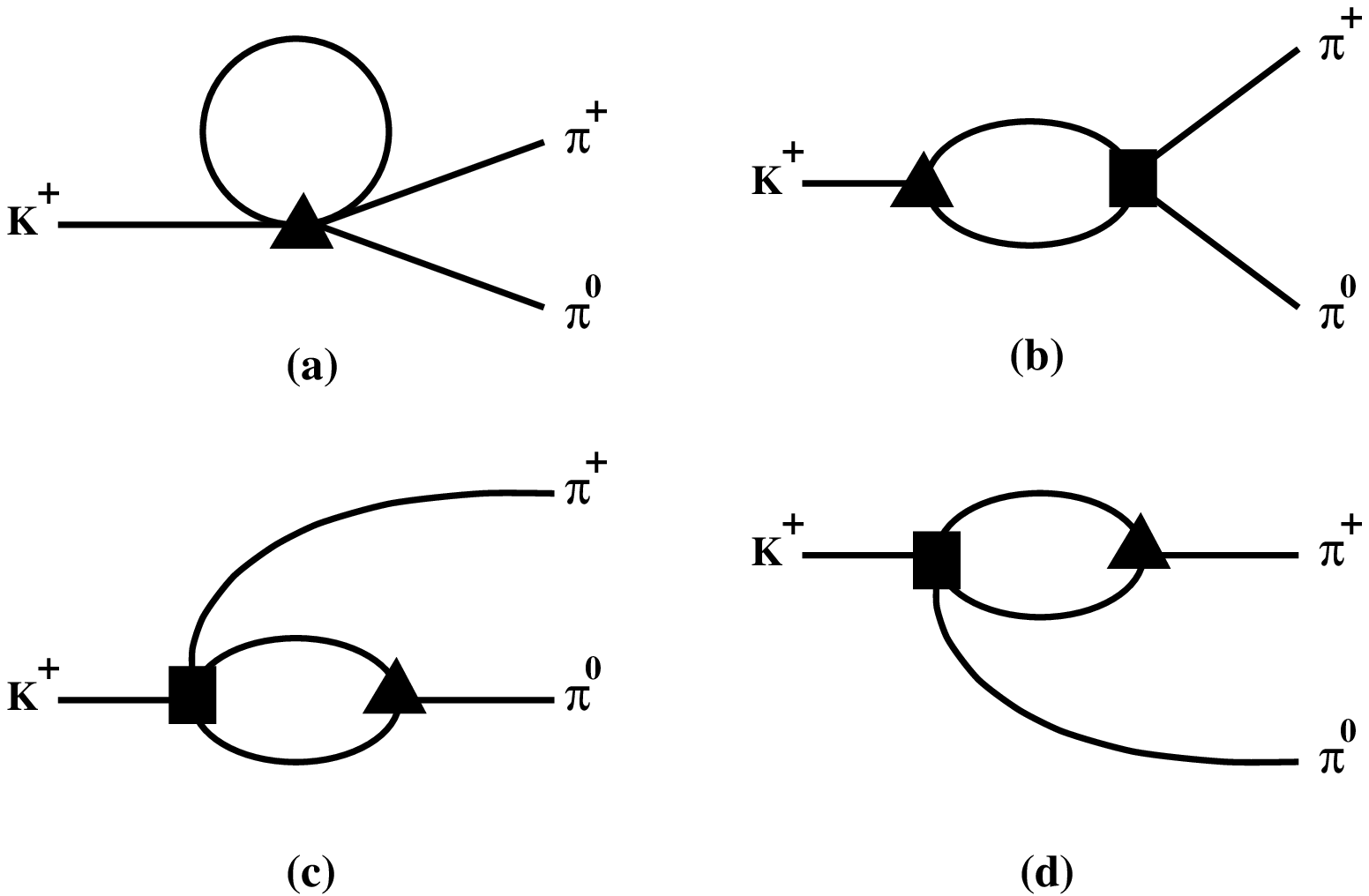}
\bigskip
{\parindent=0pt
\it Figure 1. Feynman diagrams for the one-loop contributions to the 
$K^+ \rightarrow \pi^+ \pi^0$ matrix element of $O_4$. Filled triangles 
denote weak-interaction vertices and filled squares denote 
strong-interaction vertices.}
\bigskip

We have repeated the
calculation with nonzero pion mass and $m_u =m_d \neq m_s$. Loop corrections 
come from diagrams (a) to (d) in fig. 1 and, in addition, 
from wavefunction renormalizations. Our result is: 
$$\eqalignno{
&\langle \pi^+ \pi^0 |O_4 |K^+ \rangle =\cr
&{{12i\alpha_{\scriptscriptstyle 27}} \over {\sqrt{2} f^3_\pi}}
\biggl\{ (m^2_K -m^2_\pi)\biggl[ 1+{m^2_K \over (4\pi f)^2}
\Bigl( I_z +I_f 
+{13K_1 \over 6}-{K_2\over 3}-\left( {K_1\over 6}-{7K_2\over 3}\right)y_\pi 
+G_1 +G_2 y_\pi \Bigr)\biggr]\cr
&\phantom{
{{12i\alpha_{\scriptscriptstyle 27}} \over {\sqrt{2} f^3_\pi}}
\biggl\{ (m^2_K -m^2_\pi)\biggl[ 1
}
+{m^4_K \over (4\pi f)^2}\Bigl( I_a +I_b +I_{c+d} \Bigr)
\biggr\}\ ,  &(rw)\cr
}$$
where
$$\eqalignno{
I_z &={7\over 6}\log {m^2_K \over m^2_\pi} +\left[ {1\over 3}-
{1\over 12}y_\pi \right] \log {m^2_\eta \over m^2_\pi} +{3\over 2}
\left[ 1+y_\pi \right] \log {m^2_\pi \over \Lambda^2} \; ,\cr
I_f &=-3\log {m^2_K \over m^2_\pi} -3\left[ 1+2y_\pi \right] \log {m^2_\pi 
\over \Lambda^2} \; ,\cr
I_a &=
-\left[ {13\over 3}-4y_\pi \right] \log{m^2_K \over m^2_\pi}-
\left[ {4\over 3}-y_\pi +{1\over 6}y^2_\pi\right] \log {m^2_\eta 
\over m^2_\pi}-
\left[ {17\over 3}+y_\pi -{20 \over 3}y^2_\pi \right] \log{m^2_\pi \over 
\Lambda^2}\; ,\cr
I_b &
=\left[ 1-3y_\pi +2y^2_\pi \right] \biggl(\; A(y_\pi ) +i\; \pi 
\sqrt{1-4y_\pi}\; \biggr) +
\left[ 1-{13\over 3}y_\pi +{14\over 3}y^2_\pi \right] \log{m^2_\pi \over 
\Lambda^2}\; ,\cr
I_{c+d} &=-{4\over 3y_\pi}+{4\over 3}-
\left[ {5\over 8y^2_\pi}-{13\over 8y_\pi}+1 \right] A(y_\pi )-
{1\over 72} \left[ {1\over y^2_\pi}-{1\over y_\pi} \right] B(y_\pi )\cr
&\phantom{
=-{4\over 3y_\pi}+{4\over 3}
}
+\left[ {11\over 18y^2_\pi}-{23\over 9y_\pi}+{65\over 18}-{7\over 3}y_\pi 
\right] \log {m^2_K \over m^2_\pi}\cr
&\phantom{
=-{4\over 3y_\pi}+{4\over 3}
}
+\left[ {1\over 72y^2_\pi}-{23\over 72y_\pi}+{19\over 18}-{19\over 12}y_\pi 
+{1\over 3}y^2_\pi \right] \log {m^2_\eta \over m^2_\pi} \cr
&\phantom{
=-{4\over 3y_\pi}+{4\over 3}
}
+{1\over 3}\left[ 5-8y_\pi -y^2_\pi \right] \log {m^2_\pi \over 
\Lambda^2} \; .&(Is)
}$$
The functions $A(y_\pi )$ and $B(y_\pi )$ are defined as follows:
$$\eqalignno{
A(y_\pi )&=\sqrt{1-4y_\pi}\; \log \left( {{1+\sqrt{1-4y_\pi}} \over 
{1-\sqrt{1-4y_\pi}}}\right)\ , &(A)\cr
B(y_\pi )&=\sqrt{1-44y_\pi +16y^2_\pi}\; \log \left( {{7-4y_\pi +
\sqrt{1-44y_\pi +16y^2_\pi}} \over {7-4y_\pi -\sqrt{1-44y_\pi 
+16y^2_\pi}}}\right)\ . &(fcnB)
}$$
The parameter $y_\pi$ is defined in \eq{ypi}. $I_z$ comes from wavefunction 
renormalizations (\eq{Zpifull} and \eq{ZKfull}), and $I_f$ comes from 
converting $f$ to $f_\pi$ in the tree-level contribution to \eq{rw}, using 
\eq{fpifull}. The associated $O(p^4)$ contact terms with coefficients $K_i$ 
have been combined. (Note that refs. [\cite{Kambor2,Kambor3}] keep the bare 
parameter $f$ in the tree-level term.) $I_a$ and $I_b$ are nonanalytic 
one-loop corrections from diagrams (a) and (b) respectively, while we have 
combined 
those from diagrams (c) and (d) in $I_{c+d}$.  The dimensionless
parameters $G_1$ and $G_2$ come from $O(p^4)$ weak
operators [\cite{Kambor1}].   We have again eliminated $m^2_\eta$ in favor of $m^2_\pi$ and 
$m^2_K$ using 
\eq{massrelation} except in the $\eta$ chiral logarithms. The imaginary term in 
$I_b$ 
constitutes the lowest nontrivial order phase information for $K^+ \rightarrow
\pi^+ \pi^0$ decay from ChPT.  

Analytic continuation of both $A(y_\pi )$ and $B(y_\pi )$ is understood for 
values of $y_\pi$ for which the arguments of the square roots would be
negative. Of particular interest to us is the function $B(y_\pi )$. The 
expression in \eq{fcnB} is applicable to the region extending from $y_\pi =0$ 
to the smaller of the two zeros of the argument of the square root, which is 
around 0.023. In the region between this value and the larger zero, which is 
around 2.7, the following expression should be used:
$$B(y_\pi)= -2\sqrt{-1+44y_\pi -16y^2_\pi}\; \arctan \left( 
{\sqrt{-1+44y_\pi -16y^2_\pi} \over {7-4y_\pi}} \right)\ . 
\eqno(continuedB)$$
The real-world value for $y_\pi \approx 1/13$ falls into this second region 
as opposed to the point $y_\pi =0$ which belongs to the first region.

If we do the calculation with $m_K =m_\pi$, we find that the combined 
contribution of $I_a$, $I_b$ and $I_{c+d}$ vanishes.

Setting $m_\pi=0$ in \eq{rw}, we get:
$$\eqalignno{
&\!\!\!\!\!\!\!\!\!\! \langle \pi^+ \pi^0 |O_4|K^+ \rangle =\cr
&\phantom{
\!\!\!\!\!\!\!\!\!\! \langle \pi^+ \pi^0 
}
{{12i\alpha_{\scriptscriptstyle 27} 
m^2_K} \over {\sqrt{2} f^3_\pi}}\Biggl\{ 1+{m^2_K \over (4\pi f)^2 }
\biggl( -{9\over 2}\log {m^2_K \over \Lambda^2}+i\; \pi +3\log{4\over 3} 
-{7\over 2}+{13K_1\over 6}-{K_2\over 3}+G_1 \biggr) \Biggr\}\! .
&(rwmassless)
}$$
The chiral logarithm of ref. [\cite{Bijnens}] appears, in addition to 
nonzero constant terms. The numerical significance of this will be discussed 
in section 7.

\subhead{\bf 5. $K^+ \rightarrow \pi^+ \pi^0$ matrix elements from the lattice}

$K^+ \rightarrow \pi^+ \pi^0$ decay has been studied on the lattice by 
two groups [\cite{lattdecay1,lattdecay2,lattdecay3,Claudetasi}], both of 
which used degenerate 
quark masses and the quenched approximation.  On the lattice, a kaon $K^+$ 
at rest is created at time $t=0$, using an appropriate bilinear 
quark operator. At a much later time $t_1$ a lattice version of the 
$\Delta S=1$ and $\Delta I=3/2$ weak operator in 
\eq{currentcurrent} annihilates 
the $K^+$ and creates its decay products that include $\pi^+$ and $\pi^0$. 
Finally, at a time $t_2$ much later than $t_1$, by a judicious choice of annihilation 
operators, the two pions, both at rest (see below), are picked among the 
products and 
annihilated. While the equations in this section apply to both finite and 
infinite spatial
volumes, we will focus on the torus which is the case of interest 
to lattice computations. The whole process can be expressed in terms of the 
following correlation function:
$$\eqalignno{
C(t_2 ,t_1)&\equiv \langle 0|\pi^+ (t_2) \pi^0 (t_2)O_4 (t_1)K^- (0)|0\rangle 
&(correlation0)\cr
&=\sum_{n,m} {1 \over {\langle n|n\rangle \langle m|m\rangle}}
\langle 0|e^{Ht_2} \pi^+ (0)\pi^0 (0)e^{-Ht_2}|n \rangle 
\langle n|e^{Ht_1} O_4 (0)e^{-Ht_1}|m \rangle \langle m|K^- (0)
|0 \rangle\cr &{}&(correlation)\cr
&\!\!\!\!\!\!{\buildrel{\scriptscriptstyle {t_2 \gg t_1 \gg 0}} \over 
\longrightarrow}\ \ \ 
{
{\langle 0|\pi^+ (0)\pi^0 (0)|\pi^+ \pi^0 \rangle \langle \pi^+ \pi^0 
|O_4 (0)|K^+ \rangle \langle K^+ |K^- (0)|0 \rangle} 
\over
{\langle \pi^+ \pi^0 |\pi^+ \pi^0 \rangle \langle K^+|K^+ \rangle}
}e^{-E_{2\pi} (t_2-t_1)} e^{-m_K t_1}\ , \cr
&{}&(ltlimit)\cr
}$$
where $H$ is the QCD Hamiltonian, $\pi^+ (t_2) \equiv \int d^3 {\bf x}\; 
\pi^+ ({\bf x},t_2)$ and similarly for $\pi^0 (t_2)$ and $K^- (0)$. These
fields couple to mesons with zero momentum. The operator $O_4(t_1 ) 
\equiv \int d^3 {\bf x}\; O_4 ({\bf x}, t_1 )$ couples only to states with 
zero total momentum. The state $|\pi^+ \pi^0 \rangle$ is the zero-momentum
two-pion state with lowest energy, and $|K^+ \rangle$ is a zero-momentum 
kaon state. The
sums over $n$ and $m$ run over complete sets of states. \eq{ltlimit} 
shows the leading term of \eq{correlation} in the large-time limits. Note that 
$E_{2\pi}$, which denotes the energy of two pions at rest in a finite volume,
deviates from its infinite-volume value $2m_\pi$ [\cite{italians,Luscher}]. 

We will also need the four-pion correlation function and the kaon propagator 
in order to extract the desired matrix element $\langle \pi^+ \pi^0 
|O_4 (0)|K^+ \rangle$:
$$\eqalignno{
\langle 0| \pi^+ (t_2) \pi^0 (t_2) \pi^- (t_1) \pi^0 (t_1)|0 \rangle
&={\sum_n {1\over {\langle n|n\rangle }}\langle 0| \pi^+ (t_2)\pi^0 (t_2)
|n \rangle \langle n| \pi^- (t_1)\pi^0 (t_1)|0 \rangle }\cr
&{\buildrel {t_2 \gg t_1} \over \longrightarrow}{{| \langle 0| \pi^+ (0) 
\pi^0 (0)|\pi^+ \pi^0 \rangle |}^2 \over {\langle \pi^+ \pi^0 |\pi^+ 
\pi^0 \rangle }} e^{-E_{2\pi}(t_2-t_1)}\ ,
&(pcorrel)
}$$
and
$$\eqalignno{
\langle 0|K^+ (t_1)K^- (0)|0 \rangle&=\sum_n {1\over {\langle n|n\rangle }}
\langle 0|K^+(t_1)|n \rangle \langle n|K^- (0)|0 \rangle\cr 
&{\buildrel {t_1 \gg 0} \over \longrightarrow}\;
{{| \langle K^+ |K^- (0)|0 \rangle |}^2 \over \langle K^+ 
|K^+ \rangle}e^{-m_K t_1}\ .
&(Kpro)
}$$

Some comments are in order. First, after taking the square root of the 
factors in front of the exponentials in the large-time limits in \eq{pcorrel} 
and \eq{Kpro}, we can extract the desired matrix element from \eq{ltlimit} 
only if we know the normalizations of the state vectors. We choose the 
complete set of states to consist of single-particle eigenstates and their 
tensor products. 
We use the canonical relativistic normalization for single-particle states:
$$\langle K^+ ({\bf p})|K^+ ({\bf q})\rangle =2L^3 E_K ({\bf p})
\delta_{{\bf p},{\bf q}}\ , \eqno(Knorm)$$
where $E_K ({\bf p})$ is the energy of the kaon with spatial momentum ${\bf
p}$, $L^3$ is the spatial volume, 
and $\delta_{{\bf p},{\bf q}}$ is the Kronecker delta, with similar 
definitions for other mesons. The normalizations for two-particle states 
are
$$\langle \pi^+ ({\bf P})\pi^0 ({\bf p})|\pi^+ ({\bf Q})\pi^0 ({\bf q}) 
\rangle =\langle \pi^+ ({\bf P})|\pi^+ ({\bf Q})\rangle \langle 
\pi^0 ({\bf p})|\pi^0 ({\bf q})\rangle\ .\eqno(pnorm)$$

Second, the process corresponding to the large time 
leading order contribution in 
\eq{ltlimit} does not conserve energy (except when $m_K =2m_\pi$). We
will therefore refer to the matrix element $\langle \pi^+ \pi^0 |O_4 (0)
|K^+ \rangle$ as ``unphysical." The physical matrix element corresponds to 
an excited state in \eq{correlation} if $m_K>2m_\pi$, subleading in the
large-time limits.  In principle, since in this ``unphysical" case $O_4$ 
inserts momentum, its chiral representation could 
include total-derivative terms. 
However, there are no such terms to order $p^2$. 
It is clear that we are unable to uncover the phase of any 
matrix element, in accordance with the discussion in ref. [\cite{MandT}].

\subhead{\bf 6. Calculation of $\langle \pi^+ \pi^0 |O_4 (0)|K^+ \rangle$ in
finite-volume (Q)ChPT}

While lattice computations are done at a nonzero lattice spacing (an
issue we will not address in this paper), they are also confined to a finite 
four-volume. We will take the finite spatial volume as a cube with linear 
dimension $L$ on each side. Gasser and Leutwyler, among others 
[\cite{fv}], argued that when periodic boundary conditions are used, spatial 
finite-volume effects can be taken into account in ChPT by changing the 
(spatial) momentum integrals in loop computations into discrete sums, 
{\it i.e.} for any function $f({\bf k})$, the integral $\int {{d^3 {\bf k}} 
\over (2 \pi)^3}f({\bf k})$ is changed into the sum ${1 \over L^3}\sum_{\bf k}
{f({\bf k})}$, where ${\bf k}={2\pi{\bf n}}/L$ with ${\bf n}\in 
{Z\! \! \! Z}^3$, the discrete momentum in the box. The prescription is valid 
up to corrections vanishing faster than any power of $L^{-1}$. On the other 
hand, the values of the coefficients of all $O(p^2)$ and $O(p^4)$ operators 
remain the same as in infinite volume, again up to corrections vanishing 
faster than any power of $L^{-1}$.

\bigskip
\leftline{\it A. Tree-level contribution}

We will now describe the calculation of $C(t_2,t_1)$ (\eq{correlation0}), in 
finite-volume ChPT. We will choose $t_2 > t_1 >0$. We first show the 
relatively straightforward tree-level calculation in order to set up notation 
that will facilitate the discussion of the one-loop calculation. We define 
the $\pi^+$ two-point function:
$$D_+ (x,y)=\langle \pi^+ ({\bf x}, t) \pi^- ({\bf y},s)\rangle 
\ ,\eqno(Dplus)$$
where $x=({\bf x}, t)$ and $y=({\bf y},s)$ are space-time points. The
two-point functions $D_0$ and $D_K$ for $\pi^0$ and $K^+$ are defined 
in a similar way.
Defining the sign function
$${\rm sign}(t)=\cases{\phantom{-} 1,&$t>0$ \cr\phantom{-} 0,
&$t=0$\cr -1,&$t<0$\cr}\ ,\eqno(sign)$$
and
$$\xi^0_+ =1\ \ ,\ \ \ \xi^r_+ =-\xi^l_+ =m_\pi \; {\rm sign}(t-s)\ ,
\eqno(xiplus)$$
we have for the spatial integrals of the two-point function and its 
derivatives:
$$\eqalignno{
\int d^3 {\bf x}\  D_+ (x,y)&={\xi^0_+ \over 2m_\pi} e^{-m_\pi |t-s|}\ , \cr
\int d^3 {\bf x}\  {\partial \over {\partial y_\mu}}D_+ (x,y)&=
\delta_{\mu 4}\; {\xi^r_+ \over 2m_\pi} e^{-m_\pi |t-s|}\ , &(descript)\cr
\int d^3 {\bf x}\  {\partial \over {\partial x_\mu}}D_+ (x,y)&=
\delta_{\mu 4}\; {\xi^l_+ \over 2m_\pi} e^{-m_\pi |t-s|}\ , \cr
}$$
with similar definitions of $\xi_0$'s and $\xi_K$'s associated with 
the two-point functions $D_0$ and $D_K$, respectively.

At tree level, we substitute the relevant part of $O_4 ({\bf x}_1 ,t_1 )$ 
contributing to the $K^+$ decay:
$$O_4 ({\bf x}_1 , t_1 )\rightarrow\; -{4i\alpha_{\scriptscriptstyle 27} 
\over {\sqrt{2}f^3}}
\biggl( 4\pi^- \partial^\mu \pi^0 \partial_\mu K^+ -\partial^\mu \pi^- 
\pi^0 \partial_\mu K^+ -3\partial^\mu \pi^- \partial_\mu \pi^0 K^+ \biggr) 
({\bf x}_1 , t_1 )\ , \eqno(tree1)$$
where each field is taken at the same space-time point $({\bf x}_1 , t_1 )$, 
into \eq{correlation0} and get
$$\eqalignno{
C(t_2 ,t_1)&=-{4i\alpha_{\scriptscriptstyle 27} \over {\sqrt{2}f^3}} 
\int d^3 {\bf x}_2 \int d^3 {\tilde {\bf x}}_2 \int d^3 {\bf x}_1 
\int d^3 {\bf x}_K \cr
&\biggl( 4D_+ (x_2 ,x_1 ) \partial^\mu D_0 ({\tilde x}_2 ,x_1 )
\partial_\mu D_K (x_1 ,x_K )
-\partial^\mu D_+ (x_2 ,x_1 ) D_0 ({\tilde x}_2 ,x_1 )\partial_\mu 
D_K (x_1 ,x_K )\cr
&\phantom{
\biggl( 4D_+ (x_2 ,x_1 ) \partial^\mu D_0 ({\tilde x}_2 ,x_1 )
}
-3\partial^\mu D_+ (x_2 ,x_1 ) \partial_\mu D_0 ({\tilde x}_2 ,x_1 )
D_K (x_1 ,x_K )\biggr)\ ,
&(tree2)
}$$
where $x_1 =({\bf x}_1 ,t_1 ),\ x_2 =({\bf x}_2 ,t_2 ),\ {\tilde x}_2 
=( {\tilde{\bf x}}_2 ,t_2 ),\ x_K =({\bf x}_K , 0)$ and $\partial_\mu 
={\partial \over {\partial x_{1\mu}}}\;$. After integrating over spatial 
coordinates corresponding to the external particles, {\it i.e.} ${\bf x}_2$, 
${\tilde {\bf x}}_2$ and ${\bf x}_K$, with the help of \eq{xiplus} and 
\eq{descript}, an expression independent of ${\bf x}_1$ is left. 
Integrating over ${\bf x}_1$ then gives a trivial volume factor $L^3$. 
The tree-level result is
$$\eqalignno{
C(t_2 ,t_1 )&=-{{4i\alpha_{\scriptscriptstyle 27} L^3}\over {\sqrt{2} f^3}}
{e^{-2m_\pi (t_2 -t_1 )}\over 4m^2_\pi}{e^{-m_K t_1}\over 2m_K}
\biggl( 4\xi^0_+ \xi^r_0 \xi^l_K -\xi^r_+ \xi^0_0 \xi^l_K -3\xi^r_+ \xi^r_0 
\xi^0_K \biggr)\cr
&={{12i\alpha_{\scriptscriptstyle 27} L^3}\over {\sqrt{2} f^3}}m_\pi 
(m_K +m_\pi){e^{-2m_\pi (t_2 -t_1 )}\over 4m^2_\pi}{e^{-m_K t_1}
\over 2m_K}\ .
&(tree3)
}$$
Note that to this order $E_{2\pi}=2m_\pi$. A comparison of mass prefactors 
in \eq{tree3} and \eq{rwtreeOfour} already shows the difference between 
unphysical and physical matrix elements. We will extract the unphysical 
matrix element from $C(t_2 ,t_1 )$ after the discussion of one-loop corrections. 

\bigskip
\leftline{\it B. One-loop contributions in full ChPT}

In this subsection we discuss the one-loop contributions from diagrams (a) to 
(d) (see fig. 1). 
We will from now on restrict ourselves to the mass-degenerate case.
Diagram (a) is the tadpole diagram generated by only the weak operator 
$O_4 (t_1 )$. All octet mesons appear on the loop. Diagrams (b) to (d) involve 
the $O(p^2)$ strong-interaction four-point 
vertex $({\cal L}_{int})$. These contributions are of the form
$$C(t_2 ,t_1)({\rm b,\; c,\; d})= -\langle\pi^+ (t_2)\pi^0 (t_2) 
\Bigl( \int d^3 {\bf x}_s dt_s {\cal L}_{int}({\bf x}_s , t_s )\Bigr) 
O_4 (t_1) K^- (0)\rangle_c \ , \eqno(D1)$$
where $\langle\cdots\rangle_c$ denotes all connected 
Wick contractions, and
where the relevant terms (cubic in the octet field $\phi$ (\eq{field})
of the weak operator $O_4 (t_1)$ are taken. They fall into different 
categories, represented by the different topologies of diagrams (b) to (d). 
Due to conservation of charge and isospin by the strong interaction
vertex, only $\pi^+$ and $\pi^0$ appear on the loop in diagram (b). In 
diagrams (c) and (d), a kaon and a pion or eta 
appear on the loop. We first explain in detail 
the contribution represented by diagram (b). The two external pions are 
Wick-contracted with two pions out of the four pion fields in 
${\cal L}_{int}$. The other pion fields in ${\cal L}_{int}$ are 
Wick-contracted with two pions from $O_4$. The one field left in 
$O_4$ is Wick-contracted with the external kaon. We will denote
fields contracted between $O_4$ and ${\cal L}_{int}$ by a subscript 
``$I$." For diagram (b), we end up with the product of
$$
\Biggl(\partial^\nu \pi^-_I \pi^0_I \partial_\nu K^+ - \partial^\nu 
\pi^-_I \partial_\nu \pi^0_I K^+\Biggr)({\bf x}_1 ,t_1 ) \eqno(explicitO)
$$
and
$$\eqalignno{
\Biggl( -\pi^- \pi^0 \partial^\mu \pi^+_I \partial_\mu \pi^0_I -\partial^\mu 
\pi^- \partial_\mu \pi^0 \pi^+_I \pi^0_I
&+{1\over 2}\Bigl[ \partial^\mu \pi^- \pi^0 + \pi^- \partial^\mu \pi^0 \Bigl] 
\Bigl[\partial_\mu \pi^+_I \pi^0_I +\pi^+_I \partial_\mu \pi^0_I \Bigr]\cr
&+m^2_\pi\ \pi^- \pi^0 \pi^+_I \pi^0_I \Biggr)({\bf x}_s ,t_s)\ ,
&(explicitL)
}$$
from $O_4$ and ${\cal L}_{int}$ respectively, with a prefactor 
$-8i\alpha_{\scriptscriptstyle 27}/(\sqrt{2}f^5)$. 
Fields without subscript $I$ contract with external lines. This product leads to 
a proliferation of terms with different possible combinations of derivatives 
acting on the fields, so let us represent each term 
in \eq{explicitO} and 
\eq{explicitL} symbolically by
$$\sigma ({\bf x}_1 ,t_1)\sigma^2_I ({\bf x}_1 ,t_1)\ \ {\rm and}\ \ \Phi^2 
({\bf x}_s , t_s)\Phi^2_I ({\bf x}_s , t_s)\ , \eqno(symbols)$$
respectively. The fields $\sigma$ and $\Phi$ stand for pion or kaon
fields and their derivatives. With \eq{symbols}, all terms in \eq{D1} 
corresponding to diagram (b) can be 
written as
$$\eqalignno{
&\int\!\!\! d^3\! {\bf x}_2\!\! \int\!\!\! d^3\! {\tilde {\bf x}}_2\!\! 
\int\!\!\! d^3\! {\bf x}_1\!\! \int\!\!\! d^3\! {\bf x}_K\!\! \int\!\!\! d^3\! 
{\bf x}_s\!\! \int\!\!\! dt_s\; \langle \pi^+ ({\bf x}_2 ,t_2) \Phi 
({\bf x}_s ,t_s )\rangle \langle \pi^0 ({\tilde {\bf x}}_2 ,t_2 ) \Phi 
({\bf x}_s ,t_s )\rangle\times\cr
&\phantom{
\int\!\!\! d^3\! {\bf x}_2\!\! \int\!\!\! d^3\! {\tilde {\bf x}}_2\!\! 
\int\!\!\! d^3\! {\bf x}_1\!\! \int\!\!\! d^3\! {\bf x}_K\!\! \int\!\!\! d^3\! 
{\bf x}_s\!\! \int\!\!\! dt_s\;
}
\langle \Phi_I ({\bf x}_s ,t_s)\sigma_I ({\bf x}_1 ,t_1 )\rangle \langle 
\Phi_I ({\bf x}_s ,t_s)\sigma_I ({\bf x}_1 ,t_1 )\rangle\times\cr
&\phantom{
\int\!\!\! d^3\! {\bf x}_2\!\! \int\!\!\! d^3\! {\tilde {\bf x}}_2\!\! 
\int\!\!\! d^3\! {\bf x}_1\!\! \int\!\!\! d^3\! {\bf x}_K\!\! \int\!\!\! d^3\! 
{\bf x}_s\!\! \int\!\!\! dt_s\;
}
\langle \sigma ({\bf x}_1 ,t_1 ) K^+ ({\bf x}_K ,0)\rangle\cr
&=\int\!\!\! dt_s\!\! \int\!\!\! d^3\!{\bf x}_s\!\! \int\!\!\! d^3\! 
{\bf x}_1\; {e^{-2m_\pi |t_2 -t_s |}\over 4m^2_\pi}{e^{-m_K t_1}\over 2m_K}
\xi_+ \xi_0 \xi_K \times \cr
&\phantom{
\int\!\!\! d^3\! {\bf x}_2\!\! \int\!\!\! d^3\! {\tilde {\bf x}}_2\!\! 
\int\!\!\! d^3\! {\bf x}_1\!\! \int\!\!\! d^3\! {\bf x}_K\!\! \int\!\!\! d^3\! 
{\bf x}_s\!\! \int\!\!\! dt_s\;
}
\langle \Phi_I ({\bf x}_s ,t_s)\sigma_I ({\bf x}_1 ,t_1 )\rangle \langle 
\Phi_I ({\bf x}_s ,t_s)\sigma_I ({\bf x}_1 ,t_1 )\rangle\ . &(D2)
}$$
The $\xi$-factors are chosen according to the 
prescription of \eq{descript}, and possible associated factors
$\delta_{\mu 4}$ have not been displayed explicitly. 
At this point two simplifications 
occur. First, Lorentz contractions of derivatives 
on the same internal leg do not occur 
because of the fact that they come from the simple product of two 
Lorentz invariants $O_4$ and ${\cal L}_{int}$ (see \eq{explicitO} and 
\eq{explicitL}). Second, if a derivative is 
Lorentz-contracted with another one acting on a field 
contracted with an external field, only the $4$-th component of the 
derivative survives, according to \eq{descript}. Using the momentum 
representation of the pion two-point functions, we obtain:
$$\eqalignno{
\int dt_s \int d^3 {\bf x}_s \int d^3 {\bf x}_1\; &{e^{-2m_\pi |t_2 -t_s |}
\over 4m^2_\pi}\; {e^{-m_K t_1}\over 2m_K}\; \times \cr
&\int_k \int_l \; {\cal N}\; {e^{-i\bigl({\bf k}\cdot ({\bf x}_s -{\bf x}_1)
+k_4 (t_s -t_1)\bigr)} \over (k^2 + m^2_{\pi})}{e^{-i\bigl({\bf l}\cdot 
({\bf x}_s -{\bf x}_1)+l_4 (t_s -t_1)\bigr)} \over (l^2 + m^2_{\pi})}\ ,
&(D3)
}$$
where
$${\cal N}=-{8i\alpha_{\scriptscriptstyle 27}\over {\sqrt{2} f^5}} 
\left( -m_K {\rm sign}(t_1 ) ik_4 +k^\mu l_\mu \right)
\left( k^\mu l_\mu -m_\pi {\rm sign}(t_2 -t_s )i(k_4 +l_4) \right) \ .
\eqno(N)$$
We use the abbreviation $\int_k \equiv {1\over L^3}\sum_{\bf k} 
\int {dk_4 \over 2\pi}$ with ${\bf k}=2\pi{\bf n}/L$ and ${\bf n} 
\in Z\!\!\!Z^3$.  We then integrate over ${\bf x}_1$. This results in a 
delta function setting ${\bf k}=-{\bf l}$. Physically, this corresponds to 
the fact that $O_4(t_1 )$ only couples to states with zero total momentum. 
One of the momentum sums, say over ${\bf l}$, can be performed which leads 
to an integrand independent of ${\bf x}_s$, and subsequent integration over 
${\bf x}_s$ then gives a trivial volume factor $L^3$. The result is
$$L^3\!\! \int\!\!\! dt_s {e^{-2m_\pi |t_2 -t_s |}\over 4m^2_\pi}\; {e^{-m_K 
t_1}\over 2m_K}\;  {1\over L^3}\!\!\sum_{\bf k}\!\! 
\int\!\! {dk_4 \over 2\pi}\!\! \int\!\! {dl_4 \over 2\pi} \; {\cal N}\;
{e^{-ik_4 (t_s -t_1 )} \over (k^2_4 +{\bf k}^2 +m^2_{\pi} )}{e^{-il_4 
(t_s -t_1 )} \over (l^2_4 +{\bf k}^2 +m^2_{\pi} )}\ ,
\eqno(D4)
$$
where ${\cal N}$ now takes on the form:
$${\cal N}=-{8i\alpha_{\scriptscriptstyle 27}\over {\sqrt{2} f^5}}
\left( -m_K {\rm sign}(t_1 )ik_4 +k_4 l_4 -{\bf k}^2 \right)
\left( k_4 l_4 -{\bf k}^2 -m_\pi {\rm sign}(t_2 -t_s ) i(k_4 +l_4 )
\right)\ .$$
The following formulas are needed for the evaluation of the integrals 
over $k_4$ and $l_4$ in \eq{D4}:
$$\eqalignno{
&\int {dk_4 \over {2\pi}}\int {dl_4 \over {2\pi}}\; {\cal A}\; 
{e^{-i(k_4 +l_4)\; (t_s -t_1)} \over {(k^2_4 +\omega^2_{\bf k} )
(l^2_4 +\omega^2_{\bf k} )} }\cr
&=\cases{
\int {dk_4 \over 2\pi}\delta (t_s -t_1) -\omega_{\bf k} \delta (t_s -t_1 ) 
+{\omega^2_{\bf k} \over 4} e^{-2\omega_{\bf k} |t_s -t_1|}\ ,&${\cal A}
=k^2_4 l^2_4$\cr
{\rm sign}(t_s -t_1)\Bigl( {1\over 2} \delta (t_s -t_1) -{\omega_{\bf k} 
\over 4} e^{-2\omega_{\bf k} |t_s -t_1 |}\Bigr)\ ,&${\cal A}=ik^2_4 l_4$\cr
{1\over 2\omega_{\bf k}} \delta (t_s -t_1) -{1\over 4} e^{-2\omega_{\bf k} 
|t_s -t_1 |}\ ,&${\cal A}=k^2_4$\cr
-{1\over 4}e^{-2\omega_{\bf k} |t_s -t_1 |}\ ,&${\cal A}=k_4 l_4$\cr
{1\over 4\omega_{\bf k}} {\rm sign}(t_s -t_1 ) e^{-2\omega_{\bf k} 
|t_s -t_1 |}\ ,&${\cal A}=ik_4$\cr
{1\over 4\omega^2_{\bf k}}e^{-2\omega_{\bf k}|t_s -t_1 |}\ ,&${\cal A}=1$}
&(formulas)
}$$
where 
$$\omega_{\bf k} =\sqrt{ {\bf k}^2 +m^2_\pi}\ .\eqno(omega)$$

The strong interaction can ``occur" at any time $t_s$, a fact reflected by 
the integration over $t_s$ from $-\infty$ to $+\infty$. Carrying out the 
integrating over
$t_s$ after integrating over $k_4$ and $l_4$, we finally obtain for the 
contribution from diagram (b):
$$\eqalignno{
&\! {24i\alpha_{\scriptscriptstyle 27} L^3 \over \sqrt{2}f^3}m^2_\pi \; 
{e^{-2m_\pi
(t_2 -t_1 )}\over 4m^2_\pi}{e^{-m_K t_1}\over 2m_K} \times &(D5)\cr
&\phantom{
-{8i\alpha_{\scriptscriptstyle 27}\over \sqrt{2}f^5}
}
\!\!\! 
{1\over f^2 L^3}\Biggl( -{1\over 2}(t_2 -t_1 )+{1\over 24m_\pi}-{1\over
3m^2_\pi}\sum_{\bf k} \biggl[\int {dk_4 \over 2\pi}-\omega_{\bf k} -{m^2_\pi
\over 2\omega_{\bf k}}\biggr]
\cr
&\phantom{
={24i\alpha_{\scriptscriptstyle 27} L^3 \over \sqrt{2}f^3}{1\over f^2 
L^3}\Biggl(
}
\!\!\!
-\sum_{{\bf k} \neq 0}{1\over {\bf k}^2}\biggl[{\omega^3_{\bf k} \over
{3m^2_\pi}}+{\omega_{\bf k} \over 3} -{m^2_\pi \over 6\omega_{\bf k}}\biggr]\cr
&\phantom{
={24i\alpha_{\scriptscriptstyle 27} L^3 \over \sqrt{2}f^3}{1\over f^2 
L^3}\Biggl(
}
\!\!\!
+\sum_{{\bf k}\neq 0} {1\over {\bf k}^2}\biggl[ {2\omega^2_{\bf k} \over
3m_\pi}+{\omega_{\bf k} \over 3}-{m_\pi \over 2}-{m^2_\pi \over 12\omega_{\bf
k}}+{m^3_\pi \over 12\omega^2_{\bf k}}\biggr]e^{-2(\omega_{\bf k} -m_\pi )(t_2
-t_1 )}\Biggr)\ .
}$$
We distinguished $m_K$ from $m_\pi$ in the time exponentials even though we 
consider only the case $m_K=m_\pi$. Most momentum sums
are divergent and have to be regularized. This will be discussed in 
more detail when we get to the 
final expressions including also contributions 
from diagrams (a), (c) and (d). Here we only note that
the term proportional to $\int {dk_4\over {2\pi}}$
cancels against similar terms from diagrams (c) and (d).

Of particular interest is the contribution from values of $t_s$ between 
$t_1$ and $t_2$ and with ${\bf k}=0$, which leads 
to the term proportional to 
$t_2 -t_1$. Physically, this corresponds to the subprocess in which 
the two pions created
by $O_4 (t_1 )$ are at rest and then scatter into the final pions, both also at
rest, at any time between $t_1$ and $t_2$. This term 
can be resummed into the time exponentials associated with the out-going
pions of the tree-level contribution to $C(t_2,t_1)$ in \eq{tree3}. 
(Since there are no such pion-rescattering sub-diagrams in diagrams (c) and (d)
no similar resummation occurs in those cases.) This corresponds to 
the finite-volume
correction to the energy of two pions at rest with isospin $I=2$:
$$E_{2\pi}=2m_{\pi}+{1\over 2L^3 f^2}\ .\eqno(fvmass)$$
This volume dependence agrees
with the general analysis by L\"uscher [\cite{Luscher}] of the finite-volume 
energy shifts of two-particle 
states. Here we only encounter the tree-level energy
shift because the pion-pion scattering sub-diagram is a tree-level diagram. 

The terms suppressed by a relative factor $\exp{\bigl[-2(\omega_{\bf k}^2
-m_\pi^2)
(t_2-t_1)\bigr]}$ correspond to contributions from excited states.
In fact, if $m_K >
2m_\pi$ (a mass-nondegenerate situation), \eq{correlation} indicates that the
``physical" matrix element that conserves energy-momentum is among the
nonleading terms. However, in a typical
lattice computation, $m_K < 2m_\pi$, and the finite volume is exploited to 
obtain the leading term in
the large-time limits $(t_2 \gg t_1 \gg 0)$ in which excited states are
suppressed. Then, only the unphysical matrix element is obtained. We see that
one-loop ChPT can be used in order to estimate the contamination from excited
states, as well as the difference between the physical and unphysical matrix
elements (see next section). See Appendix B for a discussion of the
extraction of the $I=2$ pion scattering length from nonleading time exponentials 
for diagram
(b) in infinite volume.

The calculation of diagrams (c) and (d) is very similar, and we will not 
discuss them here. In these cases, there are no pion-rescattering 
sub-diagrams.

\bigskip
\leftline{\it C. Calculation in QChPT} 

We now turn to the quenched calculation.
As already mentioned in the second section, QChPT has an 
unphysical particle content. Therefore, there is no reason to believe that a 
well-defined Hamiltonian exists, and none of the
manipulations of \eq{correlation} are 
justified when applied to QChPT. The large-time
behavior of \eq{ltlimit} may not be realized. 
However, a Euclidean path-integral type 
calculation of
$C(t_2,t_1)$ to one loop in QChPT similar to the one in unquenched ChPT can be 
carried out. In particular, the tree-level contribution and 
the term proportional to 
$t_2 -t_1$ that occurs in
diagram (b) are identical to those in full ChPT. Resummation of 
the latter again gives the time
exponentials for external legs as shown in \eq{ltlimit} with $E_{2\pi}$ as in
\eq{fvmass}. We will extract the matrix element 
$\langle \pi^+ \pi^0 |O_4 (0)|K^+\rangle$ from the QChPT result for $C(t_2,t_1)$ 
in the same way as in the
unquenched case. Note
that this prescription coincides with the method 
used in numerical work
[\cite{lattdecay1,lattdecay2,lattdecay3}]. 

One of the differences between ChPT and QChPT is the explicit inclusion of the
$\eta'$ field in QChPT. The $\eta'$ shows up in QChPT in two ways
[\cite{them1,Gasser}]: through arbitrary scalar functions of the $\eta'$ field 
multiplying all terms in the Lagrangian and the weak operators, and as the ninth
component of the nonet field, {\it cf.} \eq{sigmaq}.  It is easy to check that
in the case of interest, the scalar functions do not lead 
to new contributions. Furthermore, since 
the operator that represents the chiral currents is
$$\eqalignno{
\left( {L^q_\mu}\right)_{ij}
=\left(\Sigma_q \partial_\mu \Sigma_q^\dagger\right)_{ij}
&=\left( \left(e^{i \eta' / {\scriptscriptstyle \sqrt{3}}f_q}\; \Sigma \right)
\partial_\mu \left(e^{-i \eta' / {\scriptscriptstyle \sqrt{3}}f_q}\; 
\Sigma^\dagger
\right)\right)_{ij} \cr
&=\left(-i {{\partial_\mu {\eta}'} \over {\sqrt{3} f_q}}+\Sigma \partial_\mu
\Sigma^\dagger\right)_{ij}\ ,
&(current)
}
$$
$\eta'$ couplings would in general arise from diagonal current elements. 
However, since 
$$O_4 ={1\over 2}\alpha^q_{\scriptscriptstyle 27}\left(L^q_{\mu 23}  L^{q\
\mu}_{11}+L^q_{\mu 11}  L^{q\ \mu}_{23}-L^q_{\mu 23}  L^{q\ \mu}_{22}-L^q_{\mu
22}  L^{q\ \mu}_{23}+L^q_{\mu 13}  L^{q\ \mu}_{21}+L^q_{\mu 21}  L^{q\ \mu}_{13}
\right)\ , \eqno(Ofour)$$
the $\eta'$ decouples as a result of the cancellations that happen pairwise
between the first and the third and the second and the fourth terms
(for $m_u=m_d$), and the particle
content on the loops is the same as in ChPT. Since in the mass-degenerate case 
the double
pole only occurs in $\eta'$ two-point function, 
no contributions from the double pole show up here.  
As a result, the correlation function $C(t_2,t_1)$ does not have $\delta$ and
$\alpha$ dependence at one loop. Bad chiral behavior caused
by double poles is absent. 

\bigskip
\leftline{\it D. Collected results from ChPT and QChPT}

The one-loop corrected $C(t_2 ,t_1)$ consists of the tree-level 
term and the one-loop contributions of diagrams (a) to (d) 
(for wavefunction renormalizations, see below). 
Dropping contributions from 
excited states, we obtain for the full and quenched theories, in that order:
$$\eqalignno{
C^f (t_2 ,t_1 )=\;
{{24i\alpha_{\scriptscriptstyle 27} m^2_\pi L^3} \over 
\sqrt{2}f^3}&{e^{-E_{2\pi}
(t_2 -t_1)-m_K t_1} \over 8m^2_\pi m_K}\times\cr
&\!\!\Biggl( 1 
+{m^2_\pi \over 3f^2}\biggl[ {1\over {8(m_\pi L)^3}}+C(m_\pi L)+D(m_\pi 
L)\biggr]
+{m^2_\pi \over {(4\pi f)}^2}P \Biggr)&(fullc)
}$$
and
$$\eqalignno{
C^q (t_2 ,t_1 )=\;
{{24i\alpha^q_{\scriptscriptstyle 27} m^2_\pi L^3} \over
\sqrt{2}f_q^3}&{e^{-E_{2\pi} (t_2 -t_1)-m_K t_1} \over 8m^2_\pi m_K}\times\cr
&\!\!\Biggl( 1
+{m^2_\pi \over 3f^2_q}\biggl[ {1\over {8(m_\pi L)^3}}+C(m_\pi L)+{\widetilde
D}(m_\pi L)\biggr] +{m^2_\pi \over {(4\pi f_q)}^2}{\widetilde P} \Biggr) \! ,
&(quenchedc)
}$$
where
$$\eqalignno{
C(m_\pi L)&=-{1\over (m_\pi L)^3}\sum_{{\bf k} \neq 0} {1 \over {\bf
k}^2}\biggl({\omega^3_{\bf k} \over m_\pi}+m_\pi \omega_{\bf k}-{m^3_\pi \over
2\omega_{\bf k}}\biggr)\ ,\cr
D(m_\pi L)&=+{1\over (m_\pi L)^3}\sum_{\bf k}\biggl({\omega_{\bf k} \over
m_\pi}-13{m_\pi \over \omega_{\bf k}}+{3 \over 4}{m^3_\pi \over \omega^3_{\bf
k}}\biggr)\ , &(CandDs)\cr
{\widetilde D}(m_\pi L)&=+{1\over (m_\pi L)^3}\sum_{\bf k}\biggl({\omega_{\bf k}
\over m_\pi}-4{m_\pi \over \omega_{\bf k}}+{3 \over 4}{m^3_\pi \over
\omega^3_{\bf k}}\biggr) \ ,\cr
}$$
and $\omega_{\bf k}$ is defined in \eq{omega}. The contact-term coefficients $P$
and $\widetilde P$ come from $O(p^4)$ weak operators for the full and quenched
theories, respectively. Since this result does not correspond to the physical
weak
matrix element, there is no simple relation between $P$ and $G_1$ and
$G_2$ in \eq{rw}. Again, we distinguish between $m_\pi$ and $m_K$ in the
prefactor 
for illustrative purpose only. The term proportional to
$t_2 -t_1$ in \eq{D5} has been resummed using \eq{fvmass}.
    
In order to extract the unphysical matrix element from 
$C(t_2,t_1)$, we need the
four-pion correlation function \eq{pcorrel} to lowest nontrivial
order in ChPT (with $f\rightarrow f_q$ in the quenched case):
$$\langle 0| \pi^+ (t_2)\pi^0 (t_2) \pi^- (t_1)\pi^0 (t_1)|0 \rangle
={e^{-E_{2\pi}(t_2 -t_1)} \over {4m_\pi^2}}L^6 \Biggl(1+{m_\pi^2 \over
12f^2}{1 \over {(m_\pi L)}^3}\Biggr)\ .\eqno(pcorrel2)$$ 
The second term in parentheses is the lowest nontrivial 
order finite-volume correction for the two-pion wavefunction-renormalization. 
(Note that such terms
also occur in \eq{fullc} and \eq{quenchedc}). We also need the free
kaon propagator (\eq{Kpro}):
$$ \langle 0|K^+(t_1) K^-(0)|0 \rangle ={L^3 \over 2m_K} e^{-m_K t_1}\ .
\eqno(Kpro2)$$

A cutoff is assumed to regulate the divergent momentum sums in the functions 
$C$, $D$ and $\widetilde D$ defined in \eq{CandDs}. The quartic divergences 
present in the individual
terms cancel in the combinations $C+D$ and $C+{\widetilde D}$ in 
\eq{fullc} and \eq{quenchedc}, respectively. The quadratic divergence is 
independent of $m_\pi$ and can be absorbed into 
$\alpha_{\scriptscriptstyle 27}$ or 
$\alpha_{\scriptscriptstyle 27}^q$. Note that the
quadratic divergences for the full and quenched theories are not the same. This
is another indication that the coefficients 
$\alpha_{\scriptscriptstyle 27}$
and $\alpha_{\scriptscriptstyle 27}^q$ of the ChPT and 
QChPT weak operators are different. The remaining logarithmic cutoff dependence 
can be
absorbed into the coefficients $P$ and
${\widetilde P}$. 

For large $L$, we can
estimate the size of $m_\pi$ dependence of the functions $D$ and $\widetilde D$ 
by
replacing the sum ${1\over L^3}\sum_{\bf k}$ by an integral $\int {d^3{\bf k}
\over
(\; 2\pi)^3}$, with corrections vanishing faster than any power of $L^{-1}$. The
summand in the function $C$ is singular for ${\bf k}=0$, and powers of $L^{-1}$
appear
when we convert the sum to an integral. See Appendix C for details. 
In a finite volume, similar momentum sums also appear in the one-loop results 
for
the weak decay constants and wavefunction renormalizations quoted in section 2.
These sums are not singular near ${\bf k}=0$, and we can replace the sums with
integrals as for the functions $D$ and $\tilde D$ above, with
corrections vanishing faster than any power of $L^{-1}$. This leads to the
expressions given in section 2. 

When the appropriate division is carried out in order to extract the unphysical 
weak matrix element, the finite-volume two-pion wavefunction-renormalization 
corrections (terms $\propto 1/L^3$) in \eq{fullc} or \eq{quenchedc}
and \eq{pcorrel2} cancel. The matrix element thus obtained
is renormalized by multiplying it by the wavefunction-renormalization factor
$Z_\pi \sqrt{Z_K}\; $ (\eq{Zpifull} and \eq{ZKfull} for the full
theory and \eq{Zpiquenched} and \eq{ZKquenched} for the quenched theory). We 
also
make use of state normalizations \eq{Knorm} and \eq{pnorm}. The mass-degenerate 
$(m_u =m_d
=m_s)$ one-loop corrected renormalized unphysical weak matrix elements for
the full and quenched theories in a finite volume $L^3$ then are, respectively:
$$\eqalignno{
\langle \pi^+ \pi^0 &|O_4 (0)|K^+ \rangle^{f}=\cr
&{{24i \alpha_{\scriptscriptstyle
27} m^2_\pi L^3} \over {\sqrt{2}f^3}} \Biggl(1+{m^2_\pi \over (4 \pi
f)^2}\left[ -6\log{m^2_\pi \over \Lambda^2}+F(m_\pi L)+P-K_1 -K_2 \right] 
\Biggr)&(finalfull)
}$$
and
$$\eqalignno{
\langle \pi^+ \pi^0 &|O_4 (0)|K^+ \rangle^{q}=\cr
&{{24i
\alpha^q_{\scriptscriptstyle 27} m^2_\pi L^3} \over {\sqrt{2}f^3_q}}
\Biggl(1+{m^2_\pi \over (4 \pi f_q)^2}\left[ -3\log{m^2_\pi \over
\Lambda^2}+F(m_\pi L)+{\widetilde P}-{\widetilde K}\right] \Biggr)\ ,
&(finalquenched)
}$$
up to corrections vanishing faster than any power of $L^{-1}$, where the
function
$$ F(m_\pi L)={17.827 \over {m_\pi L}}+{12\pi^2 \over {(m_\pi
L)}^3}\eqno(fvcorrection)$$
is the finite-volume correction calculated in Appendix C. Note that the
same finite-volume correction occurs in both full and quenched theories.
(This is because diagram (b) is the same in the full and quenched
theories, and because only diagram (b) gives rise to 
finite-volume corrections.)
$m_\pi$ is the measured meson mass. Contact terms with coefficients $K_{1,2}$ 
and ${\widetilde K}$
come from wavefunction renormalizations. By comparing
\eq{finalfull} and \eq{finalquenched}, aside from a change from 
$\alpha_{\scriptscriptstyle 27}$ to $\alpha^q_{\scriptscriptstyle 27}$ and $f$ 
to $f_q$, we see a
substantial change to the nonanalytic one-loop corrections 
due to quenching.
If we would trade the parameter $f$ for $f_\pi$
in the tree-level term of the full theory result, the numerical factor
multiplying
the chiral logarithm would be $-15$
instead of $-6$ in \eq{finalfull}, while there would be no change for the 
quenched result since $f_q=f_\pi$ up to $O(p^4 )$ contact terms in that case. 
Because of translation 
invariance, the matrix elements of $O_4 ({\bf x}_1 , 0)$ are trivially obtained
by
omitting the factor $L^3$ from \eq{finalfull} and \eq{finalquenched}.

We end this section with a few remarks about one-loop contributions in the mass
nondegenerate case $(m_u =m_d \neq m_s )$. Due to the mixing of $\eta'$ and
$\eta$, the $\eta$ two-point function acquires a 
double pole and hence $\alpha$ and
$\delta$
dependence ($m_u =m_d$ prevents mixing of $\pi^0$ with $\eta$ and $\eta'$). 
The $\eta$ that appears on the loop of tadpole diagram (a)
contributes in a way analogous to the way $\eta$-tadpoles contribute 
to quenched $f_K$. Hence, we expect the $\delta$-dependent term to 
be singular in the chiral limit, like in \eq{fKquenched}. The contribution from 
diagram (b)
remains the same as in the mass-degenerate case, because only pions appear on 
the loop. We therefore expect that no ``enhanced finite-volume corrections" such 
as
in
the case of pion scattering [\cite{fvscatt}] at one loop. As for
diagrams (c) and (d), virtual mesons of different masses appear on the loop. In
particular, $\eta$ appears and gives rise to double poles, and hence $\alpha$ 
and
$\delta$ dependence, but no enhanced finite-volume corrections. 

\subhead{\bf 7. Numerical examples and discussion}

Little or nothing is known about the values of the low-energy constants
$F_i$, ${\widetilde F}_j$, $G_i$, $P$ and $\widetilde P$ in 
\eqs{1lfOprime,1lqOprime,rw,finalfull,finalquenched}.  (For some
phenomenological information on the $F$'s and $G$'s, see refs.
[\cite{Kambor2,Kambor3}]; for the others no information is available at all.)
Because of this situation, we will choose reasonable values of the
cutoff $\Lambda$ and ignore the contact terms in an attempt to 
estimate the size of one-loop corrections.  Values of the cutoff
we will use in the following are $770$~MeV (the mass of the $\rho$)
and $1$~GeV.  We note that in refs. [\cite{Gasser,Kambor2,Kambor3}] the value
of the $\eta$ mass was used as cutoff in a phenomenological analysis of
$O(p^4)$ contact terms.  This makes the loop diagrams artificially small, which 
therefore would most likely underestimate the size of the 
$O(p^4)$ corrections when we set the contact terms equal to zero.

The determination of the relative size of tree-level and one-loop
contributions to the kaon matrix elements is particularly sensitive to
the definition of the low-energy parameter $\alpha_{\scriptscriptstyle
27}$, especially in any scheme that has nonvanishing quadratic
divergences.  Such divergences are subtracted by renormalizing
$\alpha_{\scriptscriptstyle 27}$, and this introduces an arbitrariness
into the determination of the relative size of tree-level and one-loop
contributions. We will ignore this ambiguity in some of what follows,
and we will consider a quantity which is independent of
$\alpha_{\scriptscriptstyle 27}$ at the end of this section.
 
Apart from the fact that the $\eta'$ is light in QChPT, the same  
considerations concerning the choice of the cutoff values that apply  
in ChPT are also applicable in QChPT. Note that the cutoff in the  
quenched theory does not have to be same as that in the full theory.  
However, since non-Goldstone hadron masses in quenched QCD are roughly
the same as in full QCD, we will again take the cutoff $\Lambda_q$ for QChPT 
to be $770$~MeV or $1$~GeV. In the following numerical  
examples, we have used   
$f_\pi =132$~MeV, $m_\pi =136$~MeV, $m_K=496$~MeV, unless  
otherwise stated.  $m_\eta$ has been determined from \eq{massrelation}.

\bigskip
\leftline{\it A. $K^+ \rightarrow \pi^+ \pi^0$ decay in the real  
world}

For $m_\pi =0$, from \eq{rwmassless}, ignoring $O(p^4 )$ contact terms, at 
$\Lambda =1$~GeV  and $\Lambda =770$~MeV
we find $0.33+0.28i$ and
$0.12+0.28i$ respectively for the relative size of the one-loop
correction to the tree-level term of the $K^+\to\pi^+\pi^0$ matrix
element at $m_\pi =0$. This is smaller 
than what one would obtain from the results of
ref. [\cite{Bijnens}], which (for the real parts) would give $0.57$ and $0.35$,
respectively. This is due to the fact that in \eq{rwmassless} we kept
contributions from the nonanalytic terms that become numerical
constants after setting $m_\pi=0$. This example demonstrates that
estimates of the relative size of one-loop corrections are quite
sensitive to the choice of the values of $O(p^4)$ low-energy constants,
or, equivalently, the cutoff, for quantities involving kaons. 
In addition, this matrix element is quite sensitive to the value of
the pion mass, as one finds by substituting $m_\pi  
=136$~MeV in \eq{rw} with $O(p^4)$ contact terms ignored, which gives 
considerably larger relative one-loop contributions for the 
real parts: $0.63+0.20i$ and $0.36+0.20i$  
respectively. ($B_K$, in \eq{Bfull}, is much less sensitive to the value of 
$m_\pi$.) 

It is instructive to estimate the change in magnitude of  
$O(p^4)$ contact terms due to a change in cutoff.  This change can
be calculated from \eq{rw}, given two values of the cutoff.
For the $m_\pi  
=0$ case, a change of $\Lambda$ from $1$~GeV to $770$~MeV induces a  
change $\Delta {\widehat G}_1=2.35$. 
For the $m_\pi =136$~MeV case, one finds $\Delta({\widehat G}_1 +y_\pi 
{\widehat G_2})=2.96$. Here, we defined ${\widehat G}_1 =G_1 +{13 \over 6}K_1-{1 
\over 6}K_2$ and ${\widehat G}_2 =G_2 -{1 \over 6}K_1+{7 \over 3}K_2$.
Since the changes in ${\widehat G}_{1,2}$ are of order one, we believe that 
using
values for the cutoff of $770$~MeV and $1$~GeV gives a reasonable
estimate of the uncertainty in the size of one-loop corrections 
coming from our ignorance of the values of the $O(p^4)$ low-energy constants
[\cite{georgi}].

\bigskip
\leftline{\it B. $K^+ \rightarrow \pi^+ \pi^0$ decay on the lattice}

The physical and unphysical $K^+\to\pi^+\pi^0$ matrix elements in the full  
theory can be related up to one loop using \eq{rw} and  
\eq{finalfull}:
$$\langle \pi^+ \pi^0 |O_4 (0)|K^+\rangle_{phys} =X\; {{m^2_K  
-m^2_\pi}\over 2M^2_\pi}\; \langle \pi^+ \pi^0 |O_4  
(0)|K^+\rangle^f_{unphys}\ ,\eqno(fmr)$$
where
$$X={{1+U+{m^2_K \over (4\pi f_\pi )^2}\left[ -{5K_1 \over 6}-{K_2  
\over 6}+G_1 -\left( {K_1 \over 6}+{2K_2 \over 3}-G_2 \right)  
y_\pi\right]}\over {1+{M^2_\pi \over (4 \pi  
F_\pi)^2}\left[-6\log{M^2_\pi \over \Lambda^2}+F(M_\pi L)+P-K_1 -K_2 \right]}}\ 
, \eqno(oneloopX)$$
where the subscript ``(un)phys" refers to the (un)physical matrix
element (we imagine the unphysical matrix element to be obtained from a lattice 
computation), and the superscript ``$f$" refers to the full theory. 
\eq{fmr} relates the real-world matrix element to the
mass-degenerate unphysical matrix element as it would be computed
on the lattice with degenerate quark masses. The relation is
obtained by reconverting the factor $1/f_\pi^3$ back to $1/f^3$ in \eq{rw}. The 
one-loop corrections in the resulting expression after 
this change are those given in \eq{rw}, but without $I_f$, and now the $O(p^4)$ 
contact terms are as in the numerator of \eq{oneloopX}. The parameter $U$, then, 
is the ratio of the nonanalytic one-loop contribution and the tree-level 
contribution in \eq{rw}, leaving out $I_f$:
$$U={m^2_K \over (4\pi f_\pi )^2}\left( I_z +{m^2_K \over {m^2_K 
-m^2_\pi}}\left( I_a +I_b +I_{c+d} \right)\right)\ ,\eqno(Udef)$$
and equals $0.089$ or $-0.015$ for
$\Lambda=1$~GeV or $\Lambda=770$~MeV respectively (we will 
ignore the imaginary part of $U$, since it does not 
contribute to the magnitude of 
the amplitude to order $p^4$). 
We use upper-case
symbols $F_\pi$ and $M_\pi$ for the decay constant and meson mass
as they would be computed on the lattice, and lower-case symbols
for the real-world values of these quantities. At tree level, $X=1$.  
Note that the ratio $X$ is independent of 
$\alpha_{\scriptscriptstyle 27}$, so that it is not sensitive to
ambiguities in the renormalization of this parameter. Note also
that the $O(p^4)$ contact terms do not cancel between the numerator
and denominator of \eq{oneloopX}.

If we take values for
the pion mass and the volume typical for lattice computations, for example 
$M_\pi=0.4$~GeV, $M_\pi L=8$, and
$F_\pi=f_\pi$ (and again set the $O(p^4)$ low-energy constants equal
to zero), we find the denominator of $X$ to be about $1.8$
(leading to $X\approx 0.6$) for $\Lambda=1$~GeV, and about $1.6$ (again 
leading to $X\approx 0.6$)  
for $\Lambda=770$~MeV. 

Actual lattice computations have all been done in the quenched
approximation [\cite{lattdecay1,lattdecay2,lattdecay3}], and it is therefore 
interesting to compare the full physical and quenched unphysical 
(mass-degenerate) matrix elements directly.
In this case, we find, using \eq{rw} and \eq{finalquenched}
$${\langle \pi^+ \pi^0 |O_4 (0)|K^+ \rangle
}_{phys}=Y\;{\alpha_{\scriptscriptstyle 27}\over
\alpha^q_{\scriptscriptstyle 27}}\left({f_q \over f}\right)^3 \; 
{{m_K^2 -m_\pi^2} \over 2M^2_\pi}\; 
{\langle \pi^+ \pi^0 |O_4 (0)|K^+ \rangle
}^q_{unphys}\ , \eqno(matrixrelation)$$
with 
$$Y={{1+U+{m^2_K \over (4\pi f_\pi )^2}\left[ -{5K_1 \over 6}-{K_2
\over 6}+G_1 -\left( {K_1 \over 6}+{2K_2 \over 3}-G_2 \right)
y_\pi\right]      }\over {1+{M^2_\pi \over (4 \pi
F_\pi)^2}\left[-3\log{M^2_\pi \over \Lambda^2_q}+F(M_\pi
L)+{\widetilde P}-{\widetilde K}\right]}}\ , \eqno(oneloopY)$$
where $\Lambda_q$ is the cutoff in QChPT.  Again, at tree level
$Y=1$.  In this case there is no reason that  
$\alpha_{\scriptscriptstyle 27} =\alpha^q_{\scriptscriptstyle 27}$ (or
$f=f_q$),
since we are not comparing matrix elements in the same theory.
In particular, \eq{matrixrelation} is sensitive to renormalization
ambiguities in these parameters.

\bigskip
{\def\tablerule{\noalign{\hrule}}
\offinterlineskip
\centerline{
\vbox{\halign{%
\ \  
\hfil#\hfil\tabskip=2em&\hfil#\hfil\tabskip=2em&\hfil#\hfil\tabskip=
3.3em&\hfil#\hfil\tabskip=2.8em&\hfil#\hfil\tabskip=2.6em&\hfil#\hfil
\strut\tabskip=2.6em&\hfil#\hfil\tabskip=.5em\cr
Lattice parameters&$M^2_\pi$&$M_\pi L$&$Y^{(1)}_{\  
(1)}$&$Y^{(0.77)}_{\ (0.77)}$&$Y^{(1)}_{\ (0.77)}$&$Y^{(0.77)}_{\  
(1)}$\cr
\noalign{\vskip.1em \hrule height0pt depth1pt \vskip.5em}
$16^3 \times 25 ({\rm or}\times  
33)$&0.168&6.55&0.72&0.69&0.77&0.65\cr
$\beta =5.7$ or&0.263&8.20&0.68&0.67&0.75&0.61\cr
$1/a=1.0\ {\rm GeV}$&0.474&11.0&0.65&0.70&0.77&0.59\cr
\noalign{\vskip.5em}\tablerule\noalign{\vskip.5em}
$24^3 \times 40$&0.303&7.77&0.65&0.66&0.72&0.59\cr
$\beta =6$ or&0.428&9.24&0.63&0.67&0.74&0.57\cr
$1/a=1.7\ {\rm GeV}$&&&&&&\cr
\noalign{\vskip.5em}\tablerule
}}
}
}
\bigskip
{\parindent=0pt
\it Table 1. The factor $Y$, \eq{oneloopY}, with contact terms ignored,
at different values of $M^2_\pi$ for different combinations  
of values of the cutoff $\Lambda$ and $\Lambda_q$ for the full and quenched  
theories respectively. Lattice parameters including the four-volume $L^3  
\times T$ are shown. The superscript on $Y$ denotes $\Lambda$ in {\rm GeV};  
the subscript on $Y$ denotes $\Lambda_q$ in {\rm GeV}. $M^2_\pi$ is in ${\rm GeV}^2$. }

The tree-level relation ($Y=1$) was used in refs. 
[\cite{lattdecay2,lattdecay3,Claudetasi}] in order 
to extract the physical weak matrix  
element from the unphysical one computed on the lattice, assuming  
$\alpha_{\scriptscriptstyle 27} =\alpha^q_{\scriptscriptstyle 27}$ and $f=f_q$.  
It was found that the lattice values, corrected in this way, were at
least $50\%$ larger than the experimental value.  It is therefore
interesting to estimate the values of the one-loop correction factor
$Y$. We will again set the $O(p^4)$ low-energy constants equal to zero,
and consider two values of the cutoff, $1$~GeV and $770$~MeV, for
both the full and quenched cases.  Since these are different theories,
there is no reason that the cutoff should be chosen the same in the
numerator and denominator of \eq{oneloopY}, and we will compute four
different values of $Y$ for each $M_\pi$ corresponding to the four
possible different choices of the cutoffs in both theories.  We will
take the spread in the values of $Y$ obtained in this way as an
estimate of the uncertainty in $Y$.  Using
$F_\pi=f_\pi$, and values for $M_\pi$ and $L$ from refs. 
[\cite{Claudetasi,lattdecay3}], we report the values of $Y$ summarized in
table 1.  

In order to get an idea about the size of finite-volume effects on $Y$, 
we list in table 2  
the values of $Y$ without the finite-volume correction $F(M_\pi  
L)$ in \eq{oneloopY}.

\bigskip
{\def\tablerule{\noalign{\hrule}}
\offinterlineskip
\centerline{
\vbox{\halign{%
\ \  
\hfil#\hfil\tabskip=2em&\hfil#\hfil\tabskip=
3.3em&\hfil#\hfil\tabskip=2.8em&\hfil#\hfil\tabskip=2.6em&\hfil#\hfil
\strut\tabskip=2.6em&\hfil#\hfil\tabskip=.5em\cr
Lattice parameters&$M^2_\pi$&$Y^{(1)}_{\  
(1)}$&$Y^{(0.77)}_{\ (0.77)}$&$Y^{(1)}_{\ (0.77)}$&$Y^{(0.77)}_{\  
(1)}$\cr
\noalign{\vskip.1em \hrule height0pt depth1pt \vskip.5em}
$16^3 \times 25 ({\rm or}\times  
33)$&0.168&0.82&0.80&0.88&0.74\cr
$\beta =5.7$ or&0.263&0.79&0.80&0.88&0.71\cr
$1/a=1.0\ {\rm GeV}$&0.474&0.79&0.88&0.98&0.71\cr
\noalign{\vskip.5em}\tablerule\noalign{\vskip.5em}
$24^3 \times 40$&0.303&0.78&0.81&0.89&0.71\cr
$\beta =6$ or&0.428&0.78&0.86&0.95&0.71\cr
$1/a=1.7\ {\rm GeV}$&&&&&\cr
\noalign{\vskip.5em}\tablerule
}}
}
}
\bigskip
{\parindent=0pt
\it Table 2. The factor $Y$, \eq{oneloopY}, with both contact terms and  
finite-volume correction $F(M_\pi L)$ ignored. For further explanation, see 
the caption of table 1.}

It is evident from a comparison of tables 1 and 2 
that the finite-volume 
correction can be substantial, of order $10$--$20$~\%. Note that the 
finite-volume 
corrections do not take into account any corrections vanishing faster 
than any power of $L^{-1}$. While these are very small at 
large values of $M_\pi L$, they may be substantial for the values 
considered here [\cite{fvscatt}].
We do not address this issue further in this paper. Of course, it should be 
stressed that these estimates of the correction factor $Y$ do not
take into account the factors $\alpha_{\scriptscriptstyle 27}
/\alpha^q_{\scriptscriptstyle 27}$ and $(f_q /f)^3$ in \eq{matrixrelation}. The 
fact that
quadratic divergences differ between the full and quenched theories
({\it cf.} section 6.D) may be a hint that the ratio $\alpha_{\scriptscriptstyle 
27}
/\alpha^q_{\scriptscriptstyle 27}$ is  
not very close to one. 

The values of $Y$ in tables 1 and 2 were evaluated by first calculating the
numerator and denominator of \eq{oneloopY}, and then dividing the results.  We
could also have first 
calculated the $O(p^4)$ terms in the numerator and denominator,
and then evaluated $Y$ as one plus the difference of these $O(p^4)$ terms,
since, in order to distinguish between these two methods, a two-loop
calculation would be necessary.  Since the values for $Y$ in tables 1 and 2 
differ from one by rather substantial amounts, we would obtain significantly
different results for $Y$.  In fact, the average values of $Y$ would have been
lowered by about $10$--$20$~\% (relative to one).  This difference reflects the 
uncertainty introduced by cutting off the chiral expansion at one-loop order.

\bigskip
\epsfbox{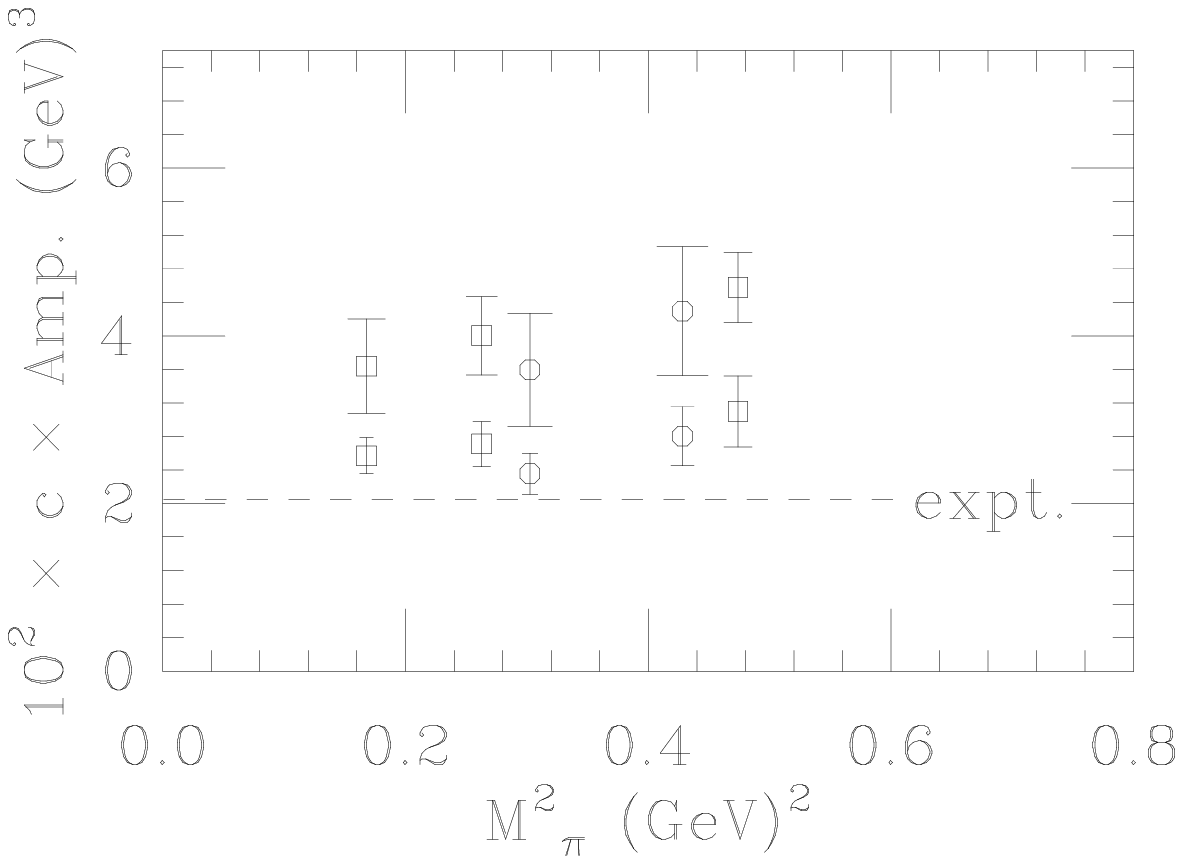}
\bigskip
{\parindent=0pt
\it Figure 2. The physical $K^+ \rightarrow \pi^+ \pi^0$ amplitude {\it vs.} 
$M^2_\pi$ from tree-level and one-loop (Q)ChPT. The one-loop corrected points 
lie below the tree-level corrected points. For detailed explanation, see text. 
Squares and octagons refer to data with four-volume $16^3 \times 25({\rm or} 
\times 33)$ and $\beta =5.7$ and four-volume $24^3 \times 40$ and $\beta =6$ in 
table 1, respectively. $c=2{\sqrt 2}/(G_F sin\theta_c cos\theta_c)$.}
\bigskip

In fig. 2, we replot some of the lattice data in fig. 16 of  
ref. [\cite{Claudetasi}] (or fig. 3 of ref. [\cite{lattdecay3}]) 
for the $K^+\to\pi^+\pi^0$ matrix element 
at different values of $M^2_\pi$ (errors are statistical). We chose data with 
$M_\pi < 770$~MeV and with the larger spatial volume for $\beta =6$. The 
$M_\pi$ and $L$ values are those used in tables 1 and 2.  
These points have already been corrected for the difference between
physical and unphysical values using tree-level ChPT,
assuming $\alpha_{\scriptscriptstyle 27}  
=\alpha^q_{\scriptscriptstyle 27}$ and 
$f=f_q$. Again assuming these equalities, we also plot the  
one-loop corrected physical matrix element using \eq{matrixrelation}  
with values of $Y$ as shown in table 1. In all cases, the one-loop corrected 
points lie below the tree-level corrected points. At each point, the central 
value 
is obtained using the average of the four different values of $Y$, while the 
error bar is obtained from the largest and smallest values of $Y$. These error 
bars do not include the statistical errors of the lattice computation. We see 
that the one-loop corrected values are closer to the 
experimental value. 

We end this sub-section with some discussion of the $O(p^4)$ low-energy
constants and the contributions from excited states to the correlation
function \eq{correlation0}.  

First, we note that, for the quenched unphysical 
weak matrix element \eq{finalquenched}, a change of cutoff  
$\Lambda_q$ from $1$~GeV to $770$~MeV results in a shift
$\Delta ({\widetilde  P}-{\widetilde K})=1.57$, again of order 
one.

Due to the fact that quenched and full QCD are different theories, 
there is no relation
between the low-energy constants 
$\widetilde P$, $\widetilde K$ and $G_{1,2}$, 
$K_{1,2}$, and therefore, in principle, no cancellation occurs in 
\eq{oneloopY}.   Alternatively, at small enough  
$M_\pi$, one may envision relating physical and unphysical weak  
matrix elements, both in the full theory, as in \eq{fmr} when  
realistic lattice computations for full QCD become  
possible. However, since the unphysical matrix element is calculated
at different external four-momenta, the contact terms $P$ for the  
unphysical full matrix element, \eq{finalfull}, and $G_{1,2}$ again  
have no simple relation, and again no cancellation takes place in 
\eq{oneloopX}. While in principle the constants $\widetilde P$,
$\widetilde K$, $P$, $K_1$ and $K_2$ can be obtained from the lattice,
this is not the case for $G_1$ and $G_2$ as long as only matrix 
elements with all mesons at rest are computed numerically.
The values of coefficients $K_1$ and $K_2$ associated with weak decay  
constants have been constrained in ref. [\cite{Gasser}] (where a different  
notation is used) with the help of large-$N_c$ arguments. A  
larger number of $O(p^4)$ operators contribute to $K^+$ decay, and 
in \eq{rw}, $G_1$ and $G_2$ are actually particular 
linear combinations of the  
associated coefficients. A determination of some of the coefficients  
in the $m_\pi =0$ approximation has been attempted in refs. 
[\cite{Kambor2,Kambor3}].  

A determination of the prefactor $(f_q /f)^3$ requires a more precise
knowledge of weak decay constants both in full and quenched theories in the 
chiral limit than appears to be available at the moment.  A fit of quenched
lattice data for $F_\pi$ in the mass-degenerate case  to \eq{fpiquenched}
would give $f_q$ and ${\widetilde K}$ [\cite{CBprivate}], but $f$, $K_1$ and
$K_2$ are only poorly known from phenomenology [\cite{Gasser}]. In addition,
nothing is known about the ratio $\alpha_{\scriptscriptstyle 27}
/\alpha^q_{\scriptscriptstyle 27}$. As long as no more detailed knowledge of 
all low-energy 
constants involved is available, we are limited to the estimates of ChPT 
one-loop effects as presented in this paper.

In order to separate the contribution of the lowest
two-pion state (for which each pion has spatial momentum 
${\bf k}=0$) 
from that of the closest excited states (for which each pion has spatial 
momentum  
$|{\bf k}|=2\pi /L$ ({\it cf.} \eq{D5})) to \eq{correlation0}, we demand that 
$$\Omega =2M_\pi t\Biggl(\sqrt{1+\Bigl({2\pi \over {M_\pi  
L}}\Bigr)^2}-1\Biggr) \gg 1. \eqno(vtcond)$$
where $t\equiv t_2-t_1$ corresponds the lower bound of the fitting range 
used in a numerical computation.  
For lattice data listed in table 1, fitting ranges in the  
four-volumes $16^3 \times 25 ({\rm or}\times 33)$ and $24^3 \times 40$  
are about $5\sim 9$ and $9\sim 13$ in lattice units, 
respectively [\cite{CBprivate}]. As an example, for the  
smallest mass $M^2_\pi =0.168$ in table 1 we find $\Omega =1.58$  
for $t=5$, and condition \eq{vtcond} is not really  
satisfied. For all mass values in table 1, $\Omega$ ranges from  
$1.04$ to $1.67$. Note that in general, at sufficiently large $M_\pi  
L$ and fixed $L$ and $t$, $\Omega$ decreases as $M_\pi$ increases.  
Since there are six excited states with $|{\bf k}|=2\pi /L$, larger  
values of $\Omega$ would be needed to be confident that the excited  
states are not contaminating the results.

\bigskip
\leftline{\it C. Relating $\langle \pi^+ \pi^0 |O_4 |K^+ \rangle$ and  
$\langle {\overline K}^0 |O'|K^0 \rangle$}

An interesting ratio is that of $\langle \pi^+  
\pi^0 |O_4 |K^+ \rangle$ and $\langle {\overline K}^0 |O'|K^0  
\rangle$, because it is independent of $\alpha_{\scriptscriptstyle 27}$ (or  
$\alpha^q_{\scriptscriptstyle 27}$ in the quenched case).  
We will consider the dimensionless quantity
$${\cal R}=f_K\; {\langle \pi^+ \pi^0 |O_4 |K^+ \rangle \over {\  
\langle {\overline K}^0 |O'|K^0 \rangle}} \eqno(R)$$
both for the physical and the unphysical case, where in the latter case
the unphysical weak matrix element $\langle \pi^+ \pi^0 |O_4 |K^+ \rangle$ is  
used. This ratio was examined in ref. [\cite{treerel}] at tree-level 
for the physical case. We can parametrize $\cal R$ as
$${\cal R}^f_{phys} ={3i\over 2\sqrt{2}}{{m^2_K -m^2_\pi}\over  
m^2_K}\left(1+R_{phys}\right)\eqno(ftreeR)$$
for the physical case, and
$${\cal R}^{f,q}_{unphys} ={3i\over  
\sqrt{2}}\left(1+R^{f,q}_{unphys}\right)\eqno(latticetreeR)$$
for the unphysical case (with degenerate quark masses). In these equations, $R$ 
stands for one-loop contributions,
{\it i.e.} $R=0$ at tree level. (Note that, at tree level there is no 
difference between the full and quenched theories in the unphysical case.)

At one loop, we evaluate $R_{phys}$ again at two choices of  
$\Lambda$: $1$~GeV and $770$~MeV, setting $O(p^4)$ contact terms equal to 
zero. 
Using \eqs{fpifull,fKfull,1lfOprime,rw} we find that 
$R_{phys}$ equals $-0.94+0.20i$ and $-0.62+0.20i$ at $\Lambda =1$~GeV and 
$770$~MeV respectively. These are large corrections, and 
call into question the reliability of ChPT for this case. 

In the unphysical (mass-degenerate) case, we find, 
from \eqs{fpifull,fKfull,1lfOprime,finalfull},
$${{{\cal R}^f_{unphys}}\over{3i/\sqrt{2}}}
=1+R^f_{unphys}=1+{M^2_\pi \over (4\pi  
F_\pi)^2}\left( 3\log{M^2_\pi \over \Lambda^2}+F(M_\pi L)+P-F_1  
+{2\over 3}\left( K_1 +K_2 \right)\right)\eqno(flatticeR)$$
for the full theory, and, from 
\eqs{fpiquenched,fKquenched,1lqOprime,finalquenched},
$${{{\cal R}^q_{unphys}}\over{3i/\sqrt{2}}}
=1+R^q_{unphys}=1+{M^2_\pi \over (4\pi  
F_\pi)^2}\left( 3\log{M^2_\pi \over \Lambda^2_q}+F(M_\pi  
L)+{\widetilde P}-{\widetilde F}_1 +{2\over 3}{\widetilde K}\right)\eqno(qlatticeR)$$
for the quenched theory. Note that the nonanalytic terms are the
same in \eq{flatticeR} and \eq{qlatticeR}.

For numerical estimates of the one-loop corrections, we  
ignore the contact terms and set $F_\pi=f_\pi$, as before. 
Table 3 lists the resulting one-loop corrections for the same 
lattice data as in table 1. Both one-loop corrections in the 
infinite-volume limit (the chiral logarithm only), labeled as  
$R^\infty$, and with finite-volume corrections ($F(M_\pi L)$), labeled  
as $R^L$, at different cutoffs are shown.

\bigskip
{\def\tablerule{\noalign{\hrule}}
\offinterlineskip
\centerline{
\vbox{\halign{%
\ \  
\hfil#\hfil\tabskip=2em&\hfil$#$\hfil\tabskip=3.3em&\hfil$#$\hfil\tabskip= 
2.3em&\hfil$#$\hfil\tabskip=2.3em&\hfil$#$\hfil\strut\tabskip=2.3em&\hfil$#$ 
\hfil\tabskip=.5em\cr
Lattice  
parameters&M^2_\pi&R^\infty_{(1)}&R^L_{(1)}&R^\infty_{(0.77)}& 
R^L_{(0.77)}\cr
\noalign{\vskip.1em \hrule height0pt depth1pt \vskip.5em}
$16^3 \times 25 ({\rm or}\times  
33)$&0.168&-0.33&-0.14\phantom{0}&-0.23&-0.039\phantom{0}\cr
$\beta =5.7$ or&0.263&-0.38&-0.15\phantom{0}&-0.23&-0.0050\cr
$1/a=1.0\ {\rm GeV}$&0.474&-0.39&-0.092&-0.12&\phantom{-}0.18\phantom{00}\cr
\noalign{\vskip.5em}\tablerule\noalign{\vskip.5em}
$24^3 \times 40$&0.303&-0.39&-0.11\phantom{0}&-0.22&\phantom{-}0.058\phantom{0}\cr
$\beta =6$ or&0.428&-0.40&-0.072&-0.15&\phantom{-}0.17\phantom{00}\cr
$1/a=1.7\ {\rm GeV}$&&&&&\cr
\noalign{\vskip.5em}\tablerule
}}
}
}
\bigskip
{\parindent=0pt
\it Table 3. Nonanalytic one-loop corrections in  
\eq{qlatticeR} in infinite volume ($R^\infty$) and in  
finite volume ($R^L$) are shown. The subscript on $R$ 
denotes $\Lambda_q$ in {\rm GeV}. $M^2_\pi$ is in ${\rm GeV}^2$.}

Large positive finite-volume   
corrections partially offset the negative corrections from chiral logarithms.  
The uncertainties for the  
ratio ${\cal R}^q_{unphys}$ associated  
with the range of cutoff explored are roughly $10\%$ to $15\%$ 
for the lower two values of $M_\pi$, and increase with $M_\pi$.

\subhead{\bf 8. Conclusion}

In this paper we have presented an analysis of the nonleptonic
kaon decay mode $K^+\to\pi^+\pi^0$ to one loop in chiral perturbation theory,
with emphasis on the comparison of results obtained from lattice QCD
computations with experiment.  One-loop ChPT makes it possible to gain analytic
insight into three different systematic errors: quenching effects, finite-volume 
effects, and effects from the fact that lattice computations (so far)
are at degenerate quark masses and have all external momenta vanishing.  The last
effect implies that on the lattice the relevant weak matrix element has been
computed at an ``unphysical" point.  We have found that, for the most recent
lattice results [\cite{lattdecay3}], all three effects are important.

Refs. [\cite{lattdecay3,Claudetasi}] already employed tree-level ChPT in order
to compare lattice results with experiment.  At tree-level no quenching or
finite-volume effects show up in ChPT, while the conversion factor between the
unphysical lattice matrix element and the physical one amounts to a simple mass
ratio (\eq{fmr} with $X=1$, or \eq{matrixrelation} with $Y=1$, $f=f_q$ and 
$\alpha_{\scriptscriptstyle 27}=\alpha^q_{\scriptscriptstyle 27}$).

This tree-level conversion factor between unphysical and physical matrix
elements gets corrections at one loop, which are embodied in the one-loop
correction factor $X$ in \eq{fmr}.  $X$ is given by \eq{oneloopX}, which also
includes finite-volume corrections.  We found that, for a meson mass
$M_\pi=0.4$~GeV and a volume $M_\pi L=8$, $X$ is approximately $0.6$, {\it
i.e.} a substantial correction.  (At $M_\pi L=\infty$, one finds 
$X\approx 0.7$.) 

\eq{matrixrelation} gives the conversion factor between the unphysical,
{\it quenched} 
matrix element in finite volume and the real-world, physical matrix
element.  The one-loop effects are contained in the factor $Y$, given in
\eq{oneloopY}.  Table 1 gives estimates of the factor $Y$ for meson masses and
spatial volumes used in ref. [\cite{lattdecay3}], while table 2 gives $Y$ for
the same masses, but in infinite volume.  From a comparison of these two
tables, we see that finite-volume effects cannot be ignored for these masses and
volumes.  In order to isolate the effect from quenching on $Y$ we can take the
ratio of $Y$ and $X$.  Considering only the chiral logarithms in this ratio, one
finds that $Y/X\approx 1.2$ for $M_\pi=0.4$~GeV, $F_\pi=f_\pi$ and
$\Lambda =\Lambda_q =1$~GeV (finite-volume corrections cancel between $Y$ and 
$X$).

In refs. [\cite{lattdecay3,Claudetasi}], lattice results for the
$K^+\to\pi^+\pi^0$ matrix element were corrected using tree-level ChPT, and
compared with experiment.  Some of the data points of ref. 
[\cite{lattdecay3,Claudetasi}] are reproduced in fig. 2.  In fig. 2 we also
showed how the lattice points would shift, if our best estimates for the
one-loop correction factor $Y$ are included (for the error bars, see section 7). 
 From fig. 2, we conclude that one-loop effects in ChPT narrow the gap
between lattice results and the experimental value.

We should stress that there are other possible
sources of discrepancy between lattice results and the experimental value. 
First, in obtaining the one-loop corrected points in fig. 2, the factors
$\alpha_{\scriptscriptstyle 27}/\alpha^q_{\scriptscriptstyle 27}$ and
$(f_q/f)^3$ in \eq{matrixrelation}
have been ignored.  Since the quenched theory is different
from the full theory, there is no {\it a priori} reason that these factors should be
equal to one.  Furthermore, our estimates of the factor $Y$ have been obtained
by choosing some values of the cutoff in (Q)ChPT, and ignoring all $O(p^4)$
low-energy constants.  While the error bars on our results in fig. 2 give some 
indication of the uncertainty introduced by this procedure, more information on 
these low-energy constants would be needed in order to be more confident about
our estimates of $Y$.  Third, since one-loop effects turn out to be rather
substantial, one may expect that two-loop corrections cannot be completely
neglected.  Also, our estimate of finite-volume effects included only
effects that go like inverse powers of $L$, and do not take into account
exponential corrections.
(For more detailed discussion of some of these points, see section
7.)  Finally, there are possible systematic errors associated with 
approximations
that we did not discuss at all, such as contamination of lattice results by
higher excited states in \eq{correlation0}, breaking of isospin symmetry and
scaling violations.  However, we believe that we have demonstrated that 
one-loop corrections in ChPT do move the lattice results closer to the 
experimental value.

If quenched lattice computations were performed with nondegenerate quark
masses, this would introduce new systematic effects into the difference between
lattice results and the experimental value of the $K^+\to\pi^+\pi^0$ matrix
element, which would show up at one loop in QChPT.  With nondegenerate masses,
the quenched result would be dependent on the ``$\eta'$-parameters" $\delta$
and $\alpha$, and presumably it would diverge in the chiral limit ({\it cf.}
section 6).  Experience with other quantities in QChPT suggests that these new
effects could be rather substantial, which would imply that nondegenerate
quenched computations may be of limited value.  

In section 4, we presented our one-loop calculation of the unquenched, physical
$K^+\to\pi^+\pi^0$ matrix element with nondegenerate masses.  For
$m_\pi=0$, our result agrees with that of ref. [\cite{Bijnens}].
However, we note that this quantity is particularly sensitive to the
pion mass, so that its numerical value at the physical pion mass can
deviate significantly from that at $m_\pi=0$.

Finally, we reviewed and extended previous existing one-loop
calculations in ChPT of the kaon B-parameter, $B_K$ 
[\cite{Bijnens,bardeen1,Kambor3,sharpe1,Bruno,sharpetasi}]
({\it cf.} section 3). We
extended the quenched calculation of refs. [\cite{sharpe1,sharpetasi}]
to include the dependence on the $\eta'$-parameter $\alpha$.
Furthermore, we discussed a modified version that could be of use for
staggered fermion QCD ({\it cf.} Appendix A).  In section 7, we used our
results in order to calculate a quantity which is basically the ratio of
the $K^0 -\Kbar^0$ and $K^+\to\pi^+\pi^0$ matrix elements.  This ratio
is independent of the $O(p^2)$ low-energy constant
$\alpha_{\scriptscriptstyle 27}$,
and therefore is of some interest (it is not independent of $O(p^4)$
low-energy constants). For the real
world, we found that this ratio has large one-loop corrections (mostly
due to large chiral logarithms in the $K^0 -\Kbar^0$ matrix element).  For the
lattice data of ref. [\cite{lattdecay3}] these one-loop corrections
change the tree-level value by some $+20$ to $-40$~\% ({\it cf.} table 3).

\subhead{\bf Acknowledgements}

We would like to thank Steve Sharpe, Akira Ukawa and, in particular, Claude
Bernard for discussions.  This work is supported in part by the US Department of
Energy (MG as an Outstanding Junior Investigator).

\subhead{\bf Appendix A: $B^q_K$ with staggered fermions}

In this Appendix, we present a somewhat modified expression for $B_K^q$
which may be relevant for quenched staggered-fermion lattice computations of 
this quantity at values of the lattice spacing where the continuum
flavor symmetry is not yet completely restored. 
It has been observed [\cite{RG,stagmeson}] that the double-pole contribution 
from
the Goldstone mesons appearing on the loops in QChPT have to be
(staggered-)flavor singlets, while the external mesons are usually 
taken to be the one exact Goldstone boson (per staggered fermion field)
of staggered-fermion QCD, which is not a flavor singlet. At typical values of 
the lattice spacing, 
one observes that the
nonexact Goldstone bosons are more or less degenerate [\cite{japmass}], while 
their
masses do not go to zero for zero quark mass, due to the staggered
breaking of chiral symmetry. Of course,
(Q)ChPT, being a continuum effective theory, is inherently inadequate to
accommodate scaling-violation effects. However, heuristically, if we denote 
the masses of the nonexact Goldstone bosons as $\overline m_K$ and
$\overline  m_\pi$,
$B_K^q$ will depend on these nonexact Goldstone masses through the
double-pole contribution to the one-loop corrections. Inserting $\overline m_K$ 
and $\overline m_\pi$ into the double-pole terms, and
ignoring $O(p^4)$ contact terms, we now get
$$\eqalignno{
B^q_K=B^q \Biggl(1
&+{m^2_K \over (4 \pi f_q)^2}\left[ -2(3+ \epsilon ^2)\log{m^2_K \over 
{\Lambda
^2}}-(2+ \epsilon ^2)\log{(1- \epsilon ^2)}-3\epsilon\log{{1+ \epsilon} 
\over {1- \epsilon}}\right]\cr
&-\delta \left[ {{\overline m}^2_K \over m^2_K}\Biggl\{-{{1- {\overline 
\epsilon}^2} 
\over {2 \overline \epsilon}}\log{{1+ \overline \epsilon} \over {1- \overline 
\epsilon}}
+1\Biggr\}-{1 \over {2 \overline \epsilon}}\log{{1+ \overline \epsilon} \over 
{1- \overline 
\epsilon}}+1 \right]\cr
&+{2 \over 3}\alpha{{\overline m}^2_K \over (4 \pi f_q)^2} \Biggl[
{{\overline m}^2_K
\over m^2_K}\Biggl\{ 2{\overline \epsilon}^2 \log{\biggl({{\overline m}^2_K 
\over 
\Lambda
^2}(1- \overline \epsilon)\biggr)}-{{1-3 {\overline \epsilon}^2-2{\overline 
\epsilon}^3} 
\over
{2 \overline \epsilon}}\log{{1+ \overline \epsilon} \over {1- \overline 
\epsilon}}+1+
{\overline \epsilon}^2\Biggr\}\cr
&
\phantom{+{2 \over 3}\alpha{{\overline m}^2_K \over (4 \pi f)^2} \Biggl[
{{\overline m}^2_K
\over m^2_K}\Biggl(2{\overline \epsilon}^2 \log{\Biggl({{\overline m}^2_K \over 
\Lambda^2}(1- \overline \epsilon)\Biggr)} } \ \
-{{1-{\overline \epsilon}^2}\over 2{\overline\epsilon}}\log{{1+{\overline 
\epsilon}} 
\over {1-{\overline \epsilon}}}+1 \Biggr]\Biggr)\ , &(modifiedBquenched)
}$$
with $\overline \epsilon= 1-{\overline m}_\pi^2/{\overline m}_K^2$. In this 
expression,
$m_K$ and $\epsilon$ refer to the exact Goldstone boson masses.
Note the appearance of terms that go like $1/{m_K^2}$.
We should stress that this expression is heuristic, since it assumes
that the only effect of nonzero lattice spacing is the appearance of
nonexact Goldstone masses in the double-pole terms. 

\subhead{\bf Appendix B: Discussion of nonleading time exponentials from 
diagram (b)}

In ref. [\cite{MandT}] Maiani and Testa showed that, for any decay
process with two identical final pions as outgoing particles, the 
leading exponential time dependence for the final pions in the corresponding 
three-point Euclidean correlation function acquires power corrections in the 
special case that both pions are at rest, and when first the 
infinite-volume limit is taken. The leading correction comes 
from the sum over all
excited states in \eq{D5} ({\it cf.} last line in that equation), which
for infinite spatial volume becomes an integral. It has the form ({\it cf.} Eq.\ 
(10) of ref.
[\cite{MandT}]) 
$$e^{-2m_\pi (t_2 -t_1)} \rightarrow e^{-2m_\pi (t_2 -t_1)}\left( 
1-a^{I=2}_0 \sqrt{{m_\pi \over {\pi (t_2 -t_1)}}}\ \right) \ .
\eqno(correction)$$
Note that the second term in \eq{correction} differs by a factor $1/2$  
as compared to Eq. (10) in ref. [\cite{MandT}]. This takes into account
the phase-space difference between distinct and identical pions in the
final state. The parameter
$a^{I=2}_0$ is the $I=2$ pion-pion $S$-wave scattering length. 
Note also that $E_{2\pi}\to 2m_\pi$ in the exponent, since this is an
infinite-volume result.  Indeed, if we replace the sum over $\bf k$ in
the last line of \eq{D5} by an integral, we find \eq{correction} with
$$a^{I=2}_0 =-{1 \over 8\pi}{m_\pi \over f^2_\pi}\ ,\eqno(scatterlength)$$
which is the tree-level ChPT expression for the $I=2$ $S$-wave scattering
length. This infinite-volume result
constitutes a useful check of our calculation.

\subhead{\bf Appendix C: Finite-volume expansion of $C( m_\pi L)$ in 
\eq{CandDs}}

In ref. [\cite{Luscher}] L\"uscher proved that, for a function $f$ of
which all derivatives are square-integrable, 
$$\sum_{{\bf k}\ne 0}{{f({\bf k}^2)}\over{({\bf k}^2)^q}}\cong L^3
{1\over{(2q-2)!}}\int {{d^3{\bf k}}\over{(2\pi)^3}}{1\over{{\bf k}^2}}
(\Delta_k)^{q-1}f({\bf k}^2)
+\sum_{j=0}^q\left({{2\pi}\over L}\right)^{2(j-q)}f^{(j)}(0)z(q-j),
\eqno(singsum)$$
with integer $q>0$ and
$f^{(j)}(0)=\left({{d\ }\over{dk^2}}\right)^jf(k^2)|_{k^2=0},$
up to corrections vanishing faster than any power of $L^{-1}$.
$\Delta_k$ is the
Laplacian with respect to ${\bf k}$ and $z(q)=Z_{00}(q,0)$ is a zeta 
function [\cite{Luscher}], with
$$
z(1)=-8.91363292\ \ ,\ \ \ z(0)=-1.\eqno(Zs)
$$

For a smooth function $g({\bf k}^2)$ for which the sum in \eq{singsum}
is ultraviolet divergent, we introduce a regulator mass $M$, and apply 
\eq{singsum} to
$$f({\bf k}^2)=g({\bf k}^2)\; \left( {M^2\over {{\bf k}^2 +M^2}}\right)^n \ ,
\eqno(modg)$$
where the positive integer $n$ is chosen sufficiently large to
make $f$ square-integrable. For $M^2\to\infty$,
$$f({\bf k}) \longrightarrow g({\bf k})\ ,\ \ (\Delta_k )^{(q-1)} f({\bf k})
\longrightarrow (\Delta_k )^{(q-1)} g({\bf k})\ \ {\rm and}\ \ f^{(j)} (0)
\longrightarrow g^{(j)} (0)\ .\eqno(Mlimit)$$
We see that the finite-volume corrections in \eq{singsum} are finite in the
$M^2\to\infty$ limit, and could have been calculated directly from $g$,
while the integral ``carries" the ultraviolet divergence.
Applying this to $C(m_\pi L)$ in \eq{CandDs} and keeping only the universal 
logarithmic cutoff dependence from the integral term, we find
$$C(m_\pi L)\to {3\over {2(4\pi)^2}}\log{m^2_\pi \over \Lambda^2}-{3z(1)\over
8\pi^2}{1\over {m_\pi L}}+{9\over 4}{1\over (m_\pi L)^3}\ ,\eqno(fvC)$$
valid up to corrections vanishing faster than any power of $L^{-1}$. 
For a discussion of the power-like ultraviolet divergences
and contact terms we refer to section 6.

\vfill
\eject

\references

\refis{phreviews}
J.F.~Donoghue, E.~Golowich and B.R.~Holstein, Dynamics of the Standard Model, 
(Cambridge, 1992); A.J.~Buras, {\it in} Proceedings of Workshop on QCD: 
20 Years 
Later, Aachen, Germany, edited by P.M.~Zerwas and H.A.~Kastrup (World 
Scientific, 1993);
E.~de~Rafael, {\it in} CP Violation and the Limits of the Standard Model, 
Proceedings of 1994 TASI School, edited by J.F.~Donoghue (World Scientific, 
1995).

\refis{BandSreview}
C.W.~Bernard and A.~Soni, {\it in} Quantum Fields on the Computer, edited by 
M.~Creutz (World Scientific, 1992).

\refis{wein}
S.~Weinberg, Phys. Rev. {\bf 166}, 1568 (1968); S.~Coleman, J.~Wess and 
B.~Zumino, Phys. Rev. {\bf 177}, 2239 (1969); C.G.~Callan, S.~Coleman, J.~Wess 
and B.~Zumino, Phys. Rev. {\bf 177}, 2247 (1969).

\refis{powercount}
S.~Weinberg, Physica {\bf 96A}, 327 (1979).

\refis{Gasser}
J.~Gasser and H.~Leutwyler, Nucl. Phys. {\bf B250}, 465 (1985).

\refis{them1}
C.W.~Bernard and M.F.L.~Golterman, Phys. Rev. {\bf D46}, 853 (1992);
Nucl. Phys. {\bf B26} (Proc. Suppl.), 360 (1992).

\refis{sharpe00}
S.R.~Sharpe, Nucl. Phys. {\bf B17} (Proc. Suppl.), 146 (1990).

\refis{sharpe0}
S.R.~Sharpe, Phys. Rev. {\bf D46}, 3146 (1992).

\refis{sharpe1}
S.R.~Sharpe, Nucl. Phys. {\bf B30} (Proc. Suppl.), 213 (1993).

\refis{fvscatt}
C.W.~Bernard and M.F.L.~Golterman, Phys. Rev. {\bf D53}, 476 (1996);
Nucl. Phys. {\bf B46}  (Proc. Suppl.), 553 (1996).

\refis{lattBK0}
R.~Gupta and T.~Bhattacharya, Nucl. Phys. {\bf B47} (Proc. Suppl.), 473 (1996).

\refis{lattBK}
D.~Pekurovsky, G.~Kilcup and L.~Venkataraman, hep-lat/9608134, 
{\it to be published in} Lattice '96, Proceedings
of the International Symposium on Lattice Field
Theory, St. Louis, U.S.A., edited by C.W.~Bernard, M.F.L.~Golterman, 
M.C.~Ogilvie and J.~Potvin; 
S.~Aoki \etal, hep-lat/9608134, {\it ibid}; 
W.~Lee and K.~Klomfass, hep-lat/9608089.

\refis{georgi}
H.~Georgi, Weak Interactions and Modern Particle Theory (Benjamin, 1984).

\refis{Kambor1}
J.~Kambor, J.~Missimer and D.~Wyler, Nucl. Phys. {\bf B346}, 17 (1990).

\refis{Kambor2}
J.~Kambor, J.~Missimer and D.~Wyler, Phys. Lett. {\bf B261}, 496 (1991).

\refis{Kambor3}
J.~Kambor, {\it in} Effective Field Theories of the Standard Model, 
Proceedings of Workshop on Effective Field Theories, 
Dobogoko, Hungary, edited by U. G. Meissner 
(World Scientific, 1992).

\refis{resonance}
G.~Ecker, J.~Gasser, A.~Pich and E.~de Rafael, Nucl. Phys. {\bf B321}, 311 
(1989).

\refis{Pichetal}
A.~Pich and E.~de Rafael, Nucl. Phys. {\bf B358}, 311 (1991); C.~Bruno and 
J.~Prades, Z.~Phys. {\bf C57}, 585 (1993). 

\refis{Bruno}
C.~Bruno, Phys. Lett. {\bf B320}, 135 (1994).

\refis{bardeen}
W.A.~Bardeen, A.J.~Buras and J.-M.~G\'erard, Phys. Lett. {\bf B192}, 138
(1987).

\refis{bardeen1}
W.A.~Bardeen, A.J.~Buras and J.-M.~G\'erard, Phys. Lett. {\bf B211}, 343 (1988).

\refis{relation}
C.W.~Bernard and M.F.L.~Golterman, Nucl. Phys. {\bf B30} (Proc. Suppl.), 217 
(1993).

\refis{zakopane}
M.F.L.~Golterman, Acta Phys. Polonica {\bf B25}, 1731 (1994).

\refis{zhang}
S.R.~Sharpe and Y.~Zhang, Phys. Rev. {\bf D53}, 5125 (1996).

\refis{sharpetasi}
S.R.~Sharpe, {\it in} CP Violation and the Limits of the 
Standard Model, Proceedings of 1994 TASI School, 
edited by J.F.~Donoghue 
(World Scientific, 1995).

\refis{treerel}
J.F.~Donoghue, E.~Golowich and B.R.~Holstein, Phys. Lett. {\bf B119}, 412 
(1982).

\refis{Bijnens}
J.~Bijnens, H.~Sonoda and M.B.~Wise, Phys. Rev. Lett. {\bf 53}, 2367 (1984).

\refis{lattdecay1}
M.B.~Gavela \etal, Nucl. Phys. {\bf B306}, 677 (1988).

\refis{lattdecay2}
C.W.~Bernard and A.~Soni, Nucl. Phys. {\bf B9} (Proc. Suppl.), 155 (1989).

\refis{lattdecay3}
C.W.~Bernard and A.~Soni, Nucl. Phys. {\bf B17} (Proc. Suppl.), 495 (1990).

\refis{MandT}
L.~Maiani and M.~Testa, Phys. Lett. {\bf B245}, 585 (1990).

\refis{fv}
J.~Gasser and H.~Leutwyler, Phys. Lett. {\bf B184}, 83 (1987); Phys. Lett. 
{\bf B188}, 477 (1987); Nucl. Phys. {\bf B307}, 763 (1988); H.~Leutwyler, 
Nucl. Phys. 
{\bf B4} (Proc. Suppl.), 248 (1988);  Phys. Lett. {\bf B189}, 197 (1987); 
H.~Neuberger, Nucl. Phys. {\bf B300}, 180 (1988); P.~Hasenfratz and 
H.~Leutwyler, Nucl. Phys. {\bf B343}, 241 (1990).

\refis{Luscher}
M.~L\"uscher, Comm. Math. Phys. {\bf 105}, 153 (1986).

\refis{sigresonance}
M.B.~Gavela \etal, Phys. Lett. {\bf B211}, 139 (1988).

\refis{isguretal}
N.~Isgur, K.~Maltman, J.~Weinstein and T.~Barnes, 
Phys. Rev. Lett. {\bf 64}, 161 
(1990).

\refis{Claudetasi}
C.W.~Bernard, {\it in} From Actions to Answers, Proceedings of the 1989 TASI 
School, edited by T.~DeGrand and D.~Toussaint (World Scientific, 1990). 

\refis{CBprivate}
C.W.~Bernard (private communication).

\refis{stagmeson}
S.R.~Sharpe, hep-lat/9609029, {\it to be published in} Lattice '96, Proceedings
of the International Symposium on Lattice Field
Theory, St. Louis, U.S.A., edited by C.W.~Bernard, M.F.L.~Golterman, 
M.C.~Ogilvie and J.~Potvin.

\refis{Booth}
M.J.~Booth, Phys. Rev. {\bf D51}, 2338 (1995); hep-lat/9412228.

\refis{italians}
H.W.~Hamber, E.~Marinari, G.~Parisi and C.~Rebbi, Nucl. Phys. {\bf B225} [FS9], 
475 (1983); G.~Parisi, Phys. Rep. {\bf 103}, 203 (1984).

\refis{RG}
R.~Gupta, Nucl. Phys. {\bf B42} (Proc. Suppl.), 85 (1995).

\refis{japmass}
N.~Ishizuka, M.~Fukugita, H.~Mino, M.~Okawa, A.~Ukawa, Nucl. Phys. {\bf B411}, 
875 (1994). 

\refis{morel}
A.~Morel,  J. Physique {\bf 48}, 111 (1987).

\endreferences

\vfill
\bye